\documentclass[a4paper, amsfonts, amssymb, amsmath, reprint, showkeys, nofootinbib, twoside]{revtex4-2}
\usepackage[english]{babel}
\usepackage{filecontents}
\usepackage{fancyhdr}
\usepackage[utf8]{inputenc}
\usepackage{amsthm}
\usepackage{mathtools}
\usepackage{physics}
\usepackage{xcolor}
\usepackage{graphicx}
\usepackage[left=13mm,right=13mm,top=25mm,bottom=25mm,columnsep=15pt]{geometry} 
\usepackage{adjustbox}
\usepackage{placeins}
\usepackage[T1]{fontenc}
\usepackage{lipsum}
\usepackage{csquotes}

\usepackage{fancyhdr} 

\fancyheadoffset{0cm}

\fancypagestyle{plain}{%
  \fancyhf{}%
  \fancyhead[R]{\thepage}%
}

\usepackage{xr-hyper}
\usepackage{upgreek}
\makeatletter

\newcommand*{\addFileDependency}[1]{
  \typeout{(#1)}
  \@addtofilelist{#1}
  \IfFileExists{#1}{}{\typeout{No file #1.}}
}
\makeatother

\newcommand*{\myexternaldocument}[1]{%
    \externaldocument{#1}%
    \addFileDependency{#1.tex}%
    \addFileDependency{#1.aux}%
}
\myexternaldocument{Supplemental}

\usepackage[pdftex, pdftitle={Article}, pdfauthor={Author}]{hyperref} 
\newcommand{\MIT}{1}
\newcommand{\USC}{2}
\newcommand{\TUB}{3}
\newcommand{\NTT}{4}

\begin{document}
\title{Deep Learning with Coherent VCSEL Neural Networks}
\author{Zaijun Chen$^{\MIT,\USC}$} 
\author{Alexander Sludds$^{\MIT}$}
\author{Ronald Davis$^{\MIT}$}
\author{Ian Christen$^{\MIT}$}
\author{Liane Bernstein$^{\MIT}$}
\author{Tobias Heuser$^{\TUB}$}
\author{Niels Heermeier$^{\TUB}$}
\author{James A. Lott$^{\TUB}$}
\author{Stephan Reitzenstein$^{\TUB}$}
\author{Ryan Hamerly$^{\MIT,\NTT}$}
\author{Dirk Englund$^{\MIT}$}

\affiliation{$^{\MIT}$Research Laboratory of Electronics, MIT, Cambridge, MA 02139, USA}
\affiliation{$^{\USC}$Ming Hsieh Department of Electrical and Computer Engineering, University of Southern California, Los Angeles, California 90089, USA}
\affiliation{$^{\TUB}$Fakultät II Institut für Festkörperphysik, Technische Universität Berlin, Berlin, 10623, Germany}
\affiliation{$^{\NTT}$NTT Research Inc., PHI Laboratories, 940 Stewart Drive, Sunnyvale, CA 94085, USA}





\begin{abstract}
  Deep neural networks (DNNs) are reshaping the field of information processing~\cite{LeCun2015}. With their exponential growth challenging existing electronic hardware~\cite{Xu2018}, optical neural networks (ONNs)~\cite{SheHar17Deep,Wright2022, Ashtiani2022, Tait2017, @LinScience2018, Feldmann2019, Feldmann2021, Xu2021, Zhou2021} are emerging to process DNN tasks in the optical domain with high clock rates, parallelism and low-loss data transmission. However, to explore the potential of ONNs, it is necessary to investigate the full-system performance incorporating the major DNN elements, including matrix algebra and nonlinear activation. Existing challenges to ONNs are high energy consumption due to low electro-optic (EO) conversion efficiency, low compute density due to large device footprint and channel crosstalk, and long latency due to the lack of inline nonlinearity~\cite{Wetzstein2020, Zhou2022, shastri2021photonicsreview}. Here we experimentally demonstrate an ONN system that simultaneously overcomes all these challenges. We exploit neuron encoding with volume-manufactured micron-scale vertical-cavity surface-emitting laser (VCSEL) transmitter arrays that exhibit high EO conversion (<5 attojoule/symbol with $V_\pi$=4 mV), high operation bandwidth (up to 25 GS/s), and compact footprint (<0.01 mm$^2$ per device). Photoelectric multiplication~\cite{Ryan_PhysRevX.9.021032} allows low-energy matrix operations at the shot-noise quantum limit. Homodyne detection-based nonlinearity enables nonlinear activation with instantaneous response. The full-system energy efficiency and compute density reach 7 femtojoules per operation (fJ/OP) and 25 TeraOP/(mm$^2\cdot$ s), both representing a >100-fold improvement over state-of-the-art digital computers~\cite{TPU_ASIC, small_footprint}, with substantially several more orders of magnitude for future improvement. Beyond neural network inference, its feature of rapid weight updating is crucial for training deep learning models. Our technique opens an avenue to large-scale optoelectronic processors to accelerate machine learning tasks from data centers to decentralized edge devices~\cite{edge_computing}.

\end{abstract}

\maketitle
\section*{Introduction} \label{sec:1}
\begin{center}
\begin{figure*}[!t]
    \centering
    \includegraphics[width=1\textwidth]{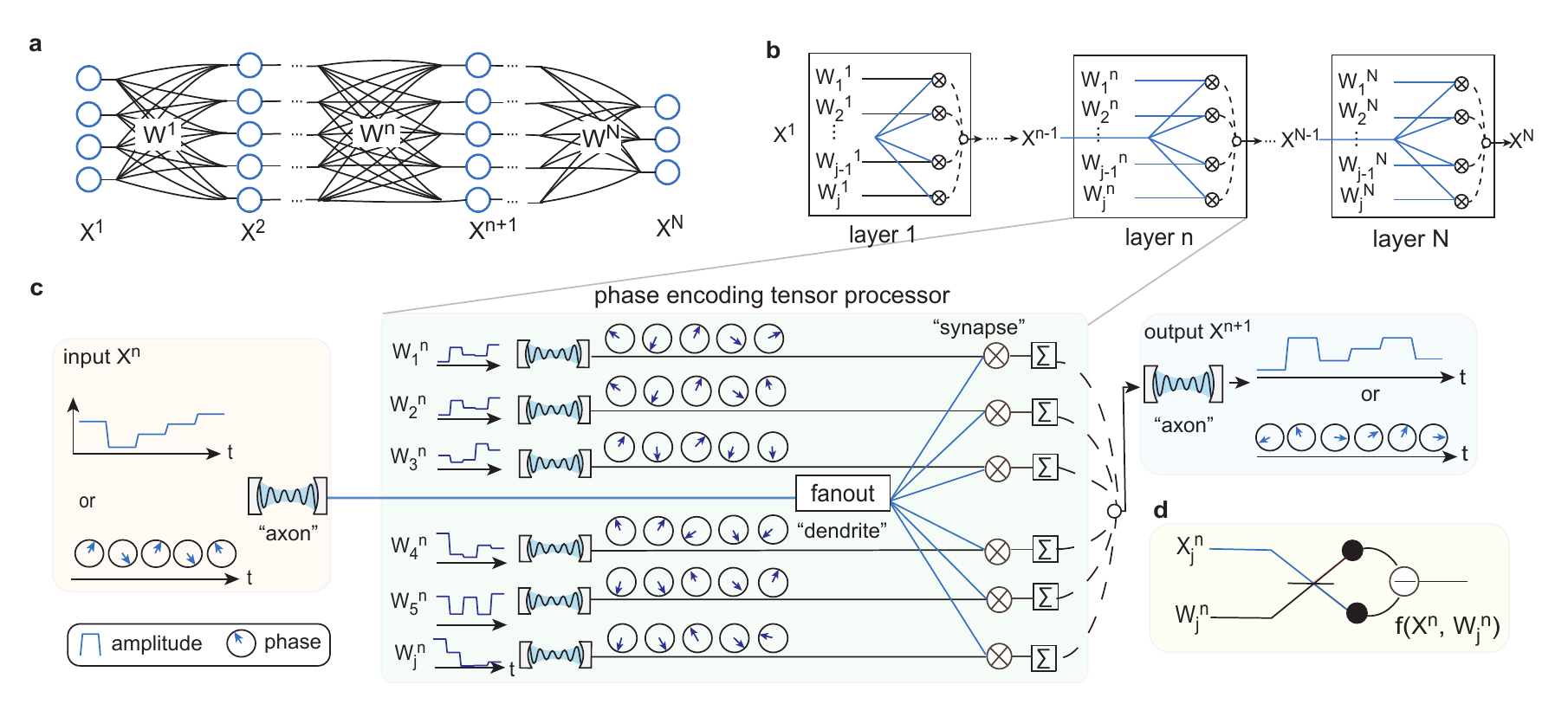}
    \caption{VCSEL-ONN architecture. \textbf{a.} Representation of fully-connected DNN composed of \textit{N} layers. \textbf{b.} Our implementation of \textit{N}-layer neural network with an optical tensor processor in each layer. \textbf{c.} Detailed illustration of the \textit{n}-th ONN layer. The ``axon'' laser oscillator encoding the input vector $X^n$ is fanned out, allowing for parallel computing with $j$ synaptic VCSELs. Multiply-and-accumulates are based on homodyne detection and time integration. The integrated products are subsequently serialized as the input vector to the next layer. \textbf{d.} Homodyne balance detection. Optical interference between two laser fields generates homodyne product $f(X^n,W_j^n$).}
    \label{fig:fig1}
\end{figure*}
\end{center}

\noindent
Artificial neural networks are computational systems that imitate the way biological brains process information. These systems are built to learn, to combine and summarize information from large data sets. Due to both advances in DNN algorithms and increases in computing power~\cite{LeCun2015}, DNNs have thrived in recent years and revolutionized information processing in applications including image~\cite{NIPS2012_c399862d}, object~\cite{object_recognition} and speech recognition~\cite{speech_process}, game playing~\cite{Mnih2015}, medicine~\cite{drug_research} and physical chemistry~\cite{molecular_dynamics}. A deep fully-connected neural network is made of $N$ layers (Fig.~\ref{fig:fig1}a), where each layer consists of a matrix-vector multiplication and a nonlinear activation. Driven by the need to tackle problems of increasing complexity, the size of machine learning models is increasing exponentially~\cite{Xu2018}, with some reaching more than 100 billion trainable parameters as of 2020~\cite{GPT3:journals/corr/abs-2005-14165}.  In contrast, due to the practical limits on transistor counts~\cite{transistor_counts} and energy consumption in data movement~\cite{datamovement}, extending computational capacity with complementary-metal–oxide–semiconductor (CMOS) circuits has become more and more difficult. An alternative approach leveraging qualitatively different technology must be developed to continue the scaling of computing power in coming decades. 

 Several critical bottlenecks emerge when designing efficient and scalable neural network accelerators. Table~\ref{tab:1} summarizes the key figures of merit, based on several recent studies~\cite{Prucnal_rev_IEEE, Wetzstein2020, Zhou2022}. State-of-the-art electronic microprocessors, such as graphics processing units (GPUs)~\cite{GPU6045685} and application-specific integrated circuits (ASICs)~\cite{TPU_ASIC}, are optimized for machine learning tasks with (C1) compute density, number of operations (OP) per second at a given chip area, $\rho=0.1$~TeraOP/(mm$^2\cdot$ s)~\cite{small_footprint} and (C2) energy efficiency, energy consumption per operation, $\epsilon \approx 1$~pJ/OP~\cite{small_footprint, TPU_ASIC}, limited by the wire capacitance of electronic interconnects~\cite{Vivienne_Tutorial}.

 \vspace{-5mm}

\begin{centering}
\begin{table}[]
    \caption{Figures of merit of VCSEL-ONN}
    \vspace{0.4cm}
    \centering
    \begin{tabular}{c c c}
       \hline
        criteria & description &  VCSEL-ONN \\
        \hline
        C1 &compute density $\rho$ & 25 TeraOP/(mm$^2\cdot s$) \\ 
        C2 & energy efficiency $\epsilon$ & 7 fJ/OP \\  
        C3 & inline nonlinearity & instantaneous response \\
        C4 & hardware scalability  & wafer-scale volume production\\
        C5 & model size  & 28$\times$28$\times$100 parameters \\
        \hline
    \end{tabular}
    \label{tab:1}
\end{table}
\end{centering}

ONNs hold great promise to alleviate this bottleneck with orders of magnitude improvement on (C1) and (C2) owing to large optical bandwidth and low-loss data transmission~\cite{miller}. Recent progress in ONNs has demonstrated fully-connected layers with photonic integrated circuits~\cite{SheHar17Deep, Tait2017, Ashtiani2022, Feldmann2019, Feldmann2021} and holographic phase masks~\cite{@LinScience2018,Zhou2021}, linear matrix operations at high throughput~\cite{Xu2021} and low-energy optical readout~\cite{sludds_netcast, wang2022optical}. However, due to the large footprint and high EO energy cost of the photonic devices (e.g., state-of-the-art low-loss EO modulators requires $V_{\pi}\cdot$$L$>1 V$\cdot$cm \cite{Wang2018_LN_modulator}), simultaneously achieving (C1) and (C2) remains an unfulfilled challenge. Moreover, incorporating low-energy all-optical nonlinearity in ONNs is challenging due to the weak photon-photon interaction. Recent advances rely on resonant cavity~\cite{Feldmann2019}, laser-cooled atoms~\cite{Zuo_Optica}, and femtosecond pulses in (millimeter-) long waveguides ~\cite{li2022allnonlinear_LN} to enhance the nonlinearity, pointing to a promising potential of speed-of-light activation, but these systems are either slow due to the cavity lifetime or limited coupling strength between atomic levels (slow rabi-oscillations), or bulky due to the device and instrument dimensions. Alternatively, most ONNs implement the nonlinear activation function digitally~\cite{SheHar17Deep, Xu2021, @LinScience2018, wang2022optical} or optoelectronically~\cite{Tait2017, PhysRevApplied.11.064043, Ashtiani2022}, resulting in latency or energy constraints. A fast, compact, and low-energy (C3) nonlinearity has yet to be developed. Furthermore, to meet the demanding scaling of DNN models, the photonic neuromorphic devices should be (C4) scalable with high density to extend computing power while reducing fabrication cost. (C5) The ONN architecture should be scalable in number of neurons to support large DNN models. State-of-the-art integrated ONNs \cite{Feldmann2021, Ashtiani2022} based on spatial multiplexing has reported models with <20 parameters.

Here, we introduce a compact VCSEL-ONN architecture, for the first time, achieves all the five criteria simultaneously. We explore the potential of high-speed VCSELs for next-generation ONNs, especially now that the VCSEL technology has matured to meet the demanding industrial requirements in 3D sensing and LiDAR~\cite{VCSEL_Lidar}, high-speed optical communications~\cite{VCSEL_review} and laser printing~\cite{laser_printing}. Our ONN system utilizes (i)~micron-scale VCSEL transceivers for high-speed (GHz) data transmission with phase coherence over the entire array via injection locking, (ii)~coherent detection for low-energy weighted accumulation, and (iii)~holographic data movement as optical dendritic fanout for parallel computing.  We experimentally achieve the-best-in-class ONN with (C1) compute density, exceeding 25 TeraOP/(s$\cdot $mm$^2$); (C2) full-system energy efficiency including digital electronics reaching 7 fJ/OP; (C3) inline nonlinearity based on homodyne detection with instantaneous response. Further, the system is (C4) scalable through existing mature wafer-scale fabrication processes and photonic integration, while high-speed (>gigahertz) time multiplexing enables the system to (C5) freely scale to run models with up to tens of billions of neurons (a model with 28$\times$28$\times$100 parameters demonstrated).

\begin{figure*}[!t]
    \centering
    \includegraphics[width=\textwidth]{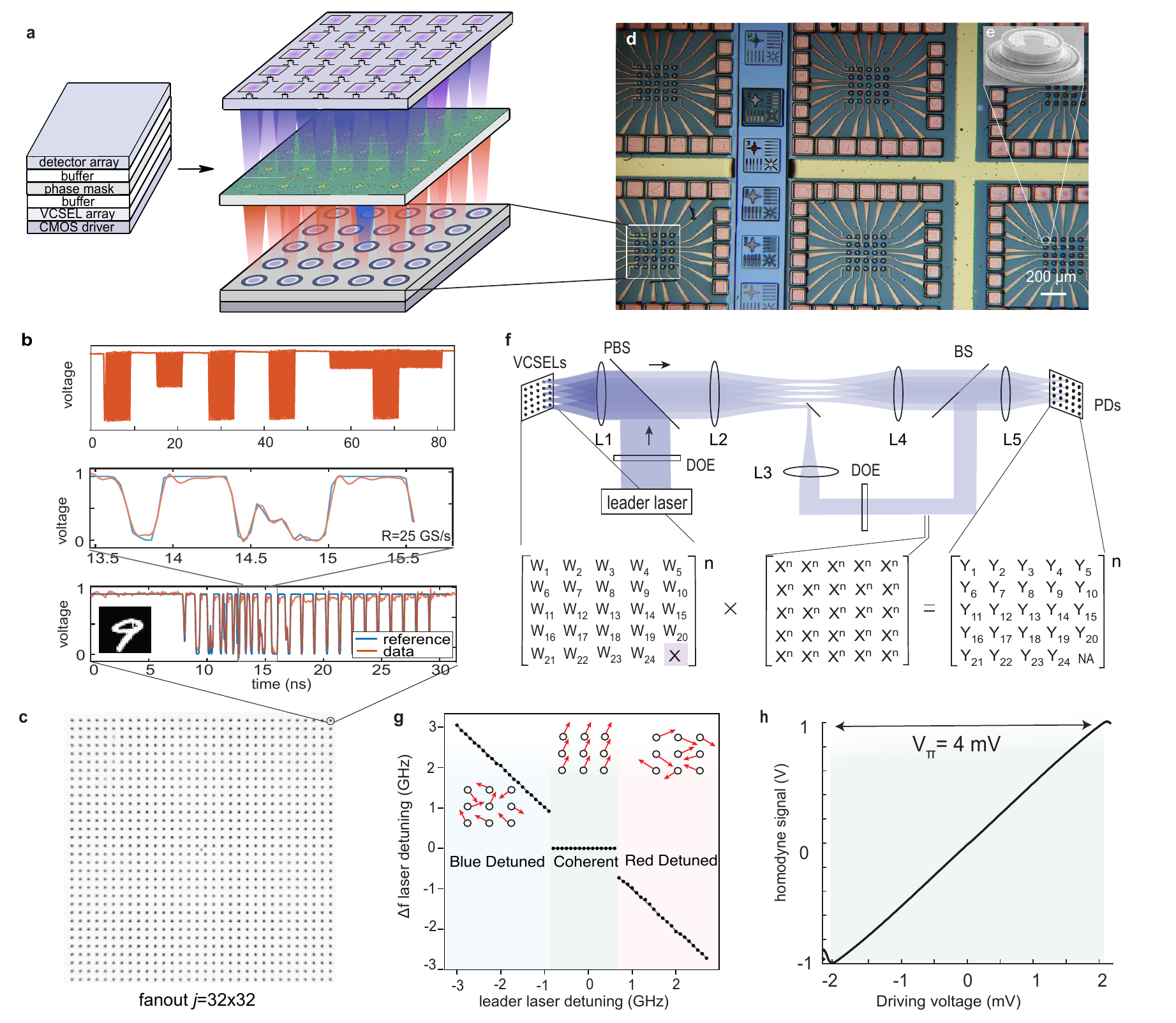}
    \caption{Experimental scheme of VCSEL-ONN. \textbf{a.} Proposed architecture with three-dimensional connectivity and photonic integration. The center (blue) is used as axon and the others (red) as weight VCSELs. The axon beam is fanned out to $j$ copies, each overlapping a weight laser onto a photo-detector.  \textbf{b.} Analog data encoding with an off-the-shelf VCSEL transmitter operating at 25 GS/s. The MIT logo in the upper plot is constructed with a time sequence of 2000 samples. An 28$\times$28-pixel image with a handwriting digit is flattened and encoded in the duration of 31.36 ns.   \textbf{c.} Demonstration of large-scale optical fanout. A single VCSEL emitting data at 25-GS/s is fannout out to 32$\times$32 spots using a diffractive optical element. A camera at the Fourier plane records the resulting beam grid. \textbf{d.} Fabricated VCSEL arrays. Arrays of 5$\times$5 wire-bonded VCSELs on a silicon substrate. \textbf{e.} Scanning electron microscopic image of a VCSEL emitter in the mist of device processing before adding the top metal connectors. \textbf{f.} Experimental setup of the VCSEL-ONN apparatus. The input data ($X^n$, highlighted) is encoded onto the corner VCSELs, while the weight matrix [$W_1^n$, ..., $W_{j}^n$] is mapped to the other VCSELs. The input VCSEL is separated from the beam arrays using a beam magnifier (L1 and L2) and D-shape mirror.  DOE: diffractive optical element. BS: beam splitter. PD: photodetector. PBS: polarizing beam splitter. \textbf{g.} Injection locking range. It is measured by monitoring the beatnote between the leader laser and each VCSEL (Supplementary Sec VI).  Within the locking range of 1.7 GHz, the VCSELs emit coherently. \textbf{h.} Low $V_\pi$ operation. Driving the VCSEL resonance over the locking range allows a phase shift of (-$\pi/2, \pi/2$), $V_\pi$=4 mV.}
    \label{fig:fig2}
\end{figure*}

\section*{schematic} \label{sec:2}
\noindent
 Our VCSEL-ONN architecture consists of a sequence of $N$ layers (Fig. \ref{fig:fig1}b). Each layer computes a matrix-vector multiplication $X^n_{(1\times i)}W^n_{(i\times j)}=Y^n_{(1\times j)}$ followed by a nonlinear activation function $f_{NL}(\cdot)$. Our scheme is similar to the ``axon-synapse-dendrite'' architecture in biological neurons. As shown in Fig. \ref{fig:fig1}c, we encode the input vector $X^n_{(1\times i)}$ in $i$ time steps to the amplitude or phase of a coherent laser oscillator (labeled ``axon''), whose beam is dendritically fanned out to $j$ copies for parallel processing. We map the weight matrix $W^n_{(i\times j)}$ in $i$ time steps with phase encoding to $j$ laser transmitters with $\sin[\phi_{W,j}(t)]\propto W_{ij}$. Each weighting laser beats with a copy of the input laser on a photo-receiver, producing the homodyne product between the two laser fields (Fig. \ref{fig:fig1}d), detailed in supplementary Sec. I. The resulting photocurrent is accumulated over $i$ time steps, yielding
 \begin{equation}\label{eq:linear_compute}
 I_j\propto\sum_i A_{W,ij}A_{X,i}\sin(\phi_{W,ij}-\phi_{X,i}), 
 \end{equation}
 
\noindent
where $A_{X,i}$ and $\phi_{X,i}$, $A_{W,ij}$ and $\phi_{W,ij}$, respectively, are the amplitude and phase of the input and weight lasers. The system allows linear multiplication $f_\textsc{L}(\cdot)\propto A_X(t)\sin[\phi_W(t)]=X_iW_{ij}$ when the input data is amplitude encoded $A_X(t)\propto X^n_i$, or nonlinear operation $f_\textsc{NL}(\cdot)\propto \sin[\phi_W(t)-\phi_X(t)]=W_{ij}\sqrt{1-X_i^2}-X_{i}\sqrt{1-W_{ij}^2}$ with phase encoding $\sin[\phi_X(t)]\propto X^n_i$. The input-output response of the two data modulation schemes is modeled (supplementary Fig.~S1), confirming linear and nonlinear operations enabled by homodyne detection.

We incorporate the detection-based optical nonlinearity in the VCSEL-ONN. As shown in Fig. S1b, programming the phase of the weight laser tunes the strength of our homodyne nonlinearity. As homodyne detection relies on the photoelectric effect, where an electron is elevated to the conduction band by the absorbed photon, the process is nearly instantaneous with a time delay of tens of attoseconds~\cite{Ossiander2018AbsoluteTO}). The resulting latency is as short as the optical pulse per symbol, which can be below femtosecond in principle. This is in contrast to the nanosecond delay with digital~\cite{SheHar17Deep, Xu2021}, electro-optic~\cite{silicon_neuron,  Tait2017} nonlinearities, and cavity- or atom-based optical nonlinearities~\cite{Zuo_Optica, Feldmann2019}. Its implementation with a photo-detector is ultra-compact without instrumental complexity (e.g., ultrashort laser pulses~\cite{li2022allnonlinear_LN}). 

 Based on space-time multiplexing and fanout data copying, the system is optimized for computing at high density and energy efficiency. It performs matrix-vector multiplication using $i$ time steps and $j$ coherent receivers. With the axon input laser shared among $j$ channels ($j$-time parallelism), the number of devices scales linearly with $O(j)$, while these requirements in the CMOS-based microprocessors~\cite{TPU_ASIC, small_footprint} and integrated ONN circuits~\cite{SheHar17Deep, @LinScience2018} scale quadratically $O(i\!\times\!j)$. The system is thus significantly simplified with reduced device counts. A constraint to the architecture is the use of one weight laser per compute channel, which gives a quadratic scaling of device counts; however, as batch operations are required in many machine learning tasks, the entire weight matrix can be spatially fanned out to $k$ copies for processing a batch of $k$ input vectors simultaneously, enabling matrix-matrix multiplication $X_{(k\times i)} W_{(i\times j)}=Y_{(k\times j)}$, with a parallelism factor of $i\times$$j$.

\begin{figure*}
    \includegraphics[width=\linewidth]{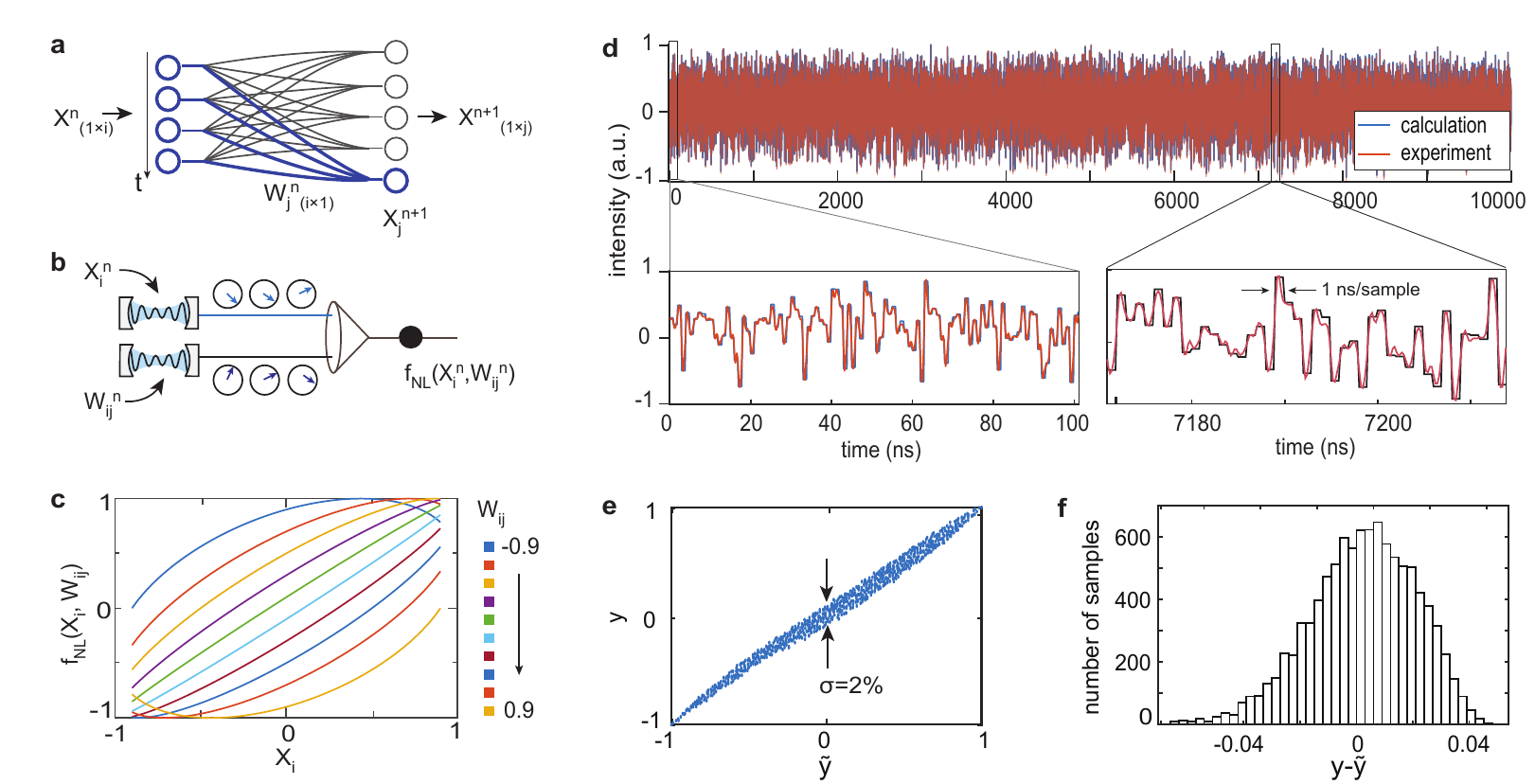}
    \caption{Characterization of compute accuracy. \textbf{a.} A neural network nonlinear compute unit, highlighted in blue. It performs vector-vector multiplication $X_j^{n+1}=\sum_i f_\textsc{NL}(X_i^n, W_{ij}^n)$ with the input vector $X^n_{1\times i}$ and the $j$-th weight vector $W_j^n {(i\times 1)}$. \textbf{b.} Each vector is phase-encoded in time steps to the output of a VCSEL. The interference between two VCSEL fields yields nonlinear weighting $f_\textsc{NL}(W_{ij}, X_i)\propto W_{ij}\sqrt{1-X_i^2}-X_{i}\sqrt{1-W_{ij}^2}$. \textbf{c.} Simulation result of homodyne interference with two phase-encoding laser fields results in nonlinear response. The strength of the nonlinearity increases at higher weights, similar to biological neural systems. \textbf{d.} Homodyne product of two normally distributed random values at the clock rate of 1 GS/s. \textbf{e.} Floating-point computation shows high accuracy with an error of less than $\sigma=2\%$. \textbf{f.} Histogram of compute errors over 10,000 input samples.}
    \label{fig:fig3}
\end{figure*}

\section*{Results}

\noindent
Our VCSEL-ONN supports a compact three-dimensional hybrid layout (Fig.~\ref{fig:fig2}a) with arrays of VCSELs for data transmission, a phase mask for beam fanout, and detector arrays for homodyne multiplication.

One of the technically challenges of this scheme is to engineer a scalable, high density, phase-stable source
of tunable, coherent outputs.  Here we exploit VCSELs for analog data encoding at rates up to 25 giga-symbol-per-second (GS/s) with 8 bits of precision: 25-billion neurons are activated in one second (Fig.~\ref{fig:fig2}b). Moreover, the VCSEL output is fanned out to 32$\times$32 copies (Fig.~\ref{fig:fig2}c) to process 1024 weights in parallel, for a potential total floating-point compute throughput of 50~TeraOP/s. However, limited by the electronic driving hardware available, we only study VCSEL arrays in size of 5$\times$5 at the operating speed of 1 GS/s. To optimize the compute density, we fabricated VCSELs based on semiconductor heterostructure microresonators with equal x- and y-direction pitches of 80 $\upmu$m (Fig. \ref{fig:fig2}d) (Methods). All the VCSELs are individually addressable with forward biasing above the lasing threshold using a battery. Each VCSEL emits 100 $\upmu$W of light with a wall-plug efficiency of 25~$\%$. The 3-dB bandwidth of our fabricated devices is $\sim$2 GHz, limited by the photon lifetime at the cavity \textit{Q}-factor of $10^{5}$ (supplementary Fig. S6). The emission wavelength of our VCSELs over the $5\times 5$ array is 974$\pm$0.1 nm. This excellent wavelength homogeneity enables parallel injection locking over the whole array to a leader laser for coherent detection (Fig. \ref{fig:fig2}f) (Methods). An injection optical power of 1~$\upmu$W per VCSEL is sufficient to achieve a stable phase lock, with a locking range of 1.7 GHz (Fig. \ref{fig:fig2}g). Tuning the individual VCSEL resonance over the locking range,  with varying driving voltages, allows phase tuning in the range $(-\pi/2, +\pi/2)$, with a $\pi$ phase shift voltage $V_{\pi}$ of 4 mV (Fig. \ref{fig:fig2}h). Such a low $V_{\pi}$ allows phase-only modulation with negligible amplitude coupling~\cite{VCSEL_modulator_Hoghooghi:10} and negligible crosstalk between neighbouring channels. 

For matrix operations, we use a single VCSEL array to encode both the input activation and weights; sharing beam paths improves the interferometric stability in homodyne detection. Among the 25 VCSELs, 24 encode weights, while the corner VCSEL encodes the activations (Fig. \ref{fig:fig2}f). Limited by the large dimension of our DOE, the corner laser is separated from the main beams and then fanned out to $9\times 9$ spots. We note that the beam separation is not necessary in the future with a compact DOE. Each fanout beam spot is superimposed to a weight laser beam $W_j^n$ with a beam splitter. A total of 5$\times$5 combined beams are coupled to a fiber-based detector array for coherent detection.

We characterize the computing accuracy of homodyne interference in our neural network implementation (Fig. \ref{fig:fig3}). We utilize two injection-locked VCSELs to construct a compute unit of vector-vector multiplication based on our physical system's unique nonlinearity $f_\textsc{NL}(X_i,W_{ij})$ (Fig. \ref{fig:fig3}c), which differs from the conventional multiplication \cite{Ryan_PhysRevX.9.021032} due to the complex-valued nature of the VCSEL network’s outputs.  We encode two vectors $X^n$ and $W_j^n$, each with $i$=10,000 normally distributed random values to a VCSEL at the clock rate of R=1 GS/s with a peak-to-peak voltage of 4 mV. The signals are AC coupled to remove the slow thermal drifts (methods). The experimental time trace agrees well with the calculation in Fig. \ref{fig:fig3}d. The standard deviation of $y-\widetilde{y}$ residuals in Fig. \ref{fig:fig3}e and f reveals a computing accuracy of 98$\%$ ($\sim$6 bits of precision), limited mainly by the phase instability of the setup and the frequency response of injection locked VCSELs. The accuracy can be improved in the future with better photonic integration and VCSELs of higher bandwidth, though this accuracy is sufficient for a wide range of machine learning tasks \cite{quantized_low_precision}.

\begin{figure*}
    \centering
    \includegraphics[width=\textwidth]{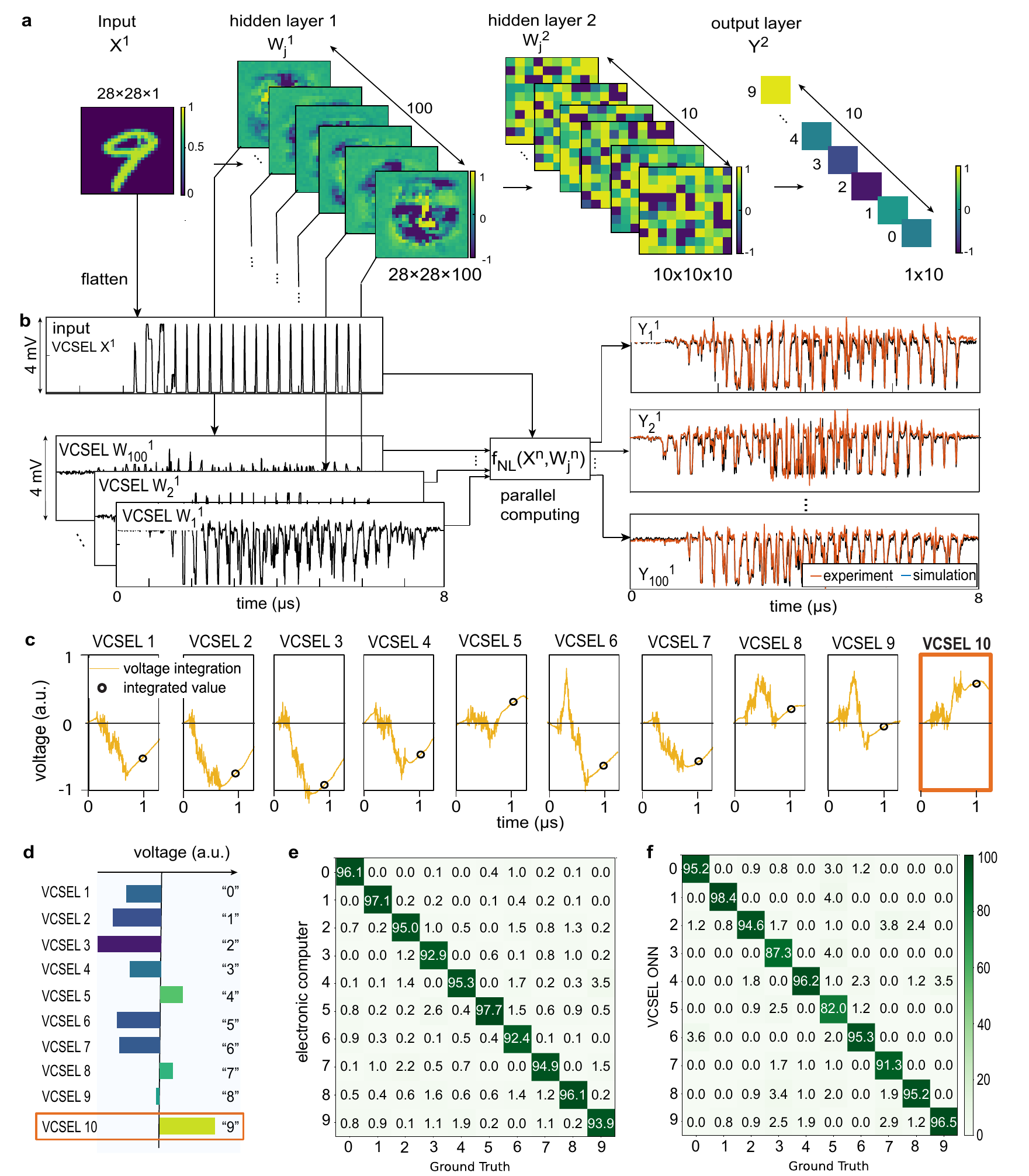}
    \caption{Benchmarking of machine learning inference with VCSEL ONN. \textbf{a.} Model for MNIST image classification trained with our unique nonlinearity $f_\textsc{{NL}}(\cdot)$. It consists of one input layer, two hidden layers and an output layer. There are 100 and 10 neurons, respectively, in the first and second hidden layers. \textbf{b.} Example parallel multiplication. The input image in layer 1 is flattened and encoded in time steps to the phase of the $X^n$ VCSEL. The weight matrix with 100 vectors is encoded to 100 individual weighting VCSELs. The result of matrix-vector multiplication (red) is compared to the theoretical signal (black). \textbf{c.} Example time integration of interference signal in layer 2. The black dots are the integrated results. \textbf{d.} Output layer.    The result of MNIST classification is read out by comparing the integrated voltage (black circles in \textbf{c}) of the 10 processing VCSEL channels.  \textbf{e.} Comparison of electrical and optical confusion matrices, showing good agreement between the ONN and the von Neumann electronic computer systems.}
    \label{fig:fig4}
\end{figure*}

We deploy a neural network inference on our VCSEL ONN to classify hand-written digits in the Modified National Institute of Standards and Technology database (MNIST).  To this end, we developed a training algorithm with PyTorch using our unique nonlinear weighting function $f_\textsc{{NL}}(W_j^n, X^n)$. Fig.~\ref{fig:fig4}a shows the trained 3-layer model (size 28$\times$28$\rightarrow$100$\rightarrow$10$\rightarrow$10). Each image, with $28\times 28$ pixels, is flattened and encoded in 784 time steps to an input VCSEL (Fig. \ref{fig:fig4}b) at the diving voltage of 4 mV. The 100 weight vectors in the first hidden layer are flattened and encoded one weight vector per VCSEL. Ideal spatial multiplexing allows processing of 100 weight channels simultaneously; however, limited by the arbitrary waveform generator (AWG) hardware, the data is taken with multiple acquisitions (five VCSELs per acquisition) at the clock rate of 100 MS/s, although the VCSEL bandwidth allows homodyne interference at > 1 GS/s (Fig.~\ref{fig:fig3}c). By switching the AWG channels and translating the VCSEL chips to different arrays in the x- and y-directions, a total 100 VCSEL devices from 5 arrays are used to compute in the first layer.  The interference signal between the image data and a weight vector is compared to the digitally-calculated result in Fig.~\ref{fig:fig4}b. We obtained $\sim$6 bits of compute precision, similar to the result in Fig.~\ref{fig:fig3}b. The signal-to-noise ratio (SNR) in the time trace is 135, limited by the photon shot noise (Supplementary Fig. S2). The photocurrent at each channel is accumulated over time with an custom-made time integrator (methods). The integrated values from the 100 channels are serialized, forming an input vector feeding into the second hidden layer. The weights in the second hidden layer is implemented with 10 weighting VCSELs and the interference signal is integrated. Fig.~\ref{fig:fig4}c shows the real-time integration of the interference signal for processing an image in layer 2. The image classification is read out by the max integrating voltage of the 10 VCSEL channels (Fig. \ref{fig:fig4}c and d). Running inference over a dataset of 1000 MNIST test images, with a total of 158.8-million operations, we obtain an accuracy of $(93.1 \pm 2)\%$, which is 98 $\%$ of the model's accuracy in simulation (95.1$\%$). 

\section*{System performance}
\begin{figure}[h]
    \includegraphics[width=0.5\textwidth]{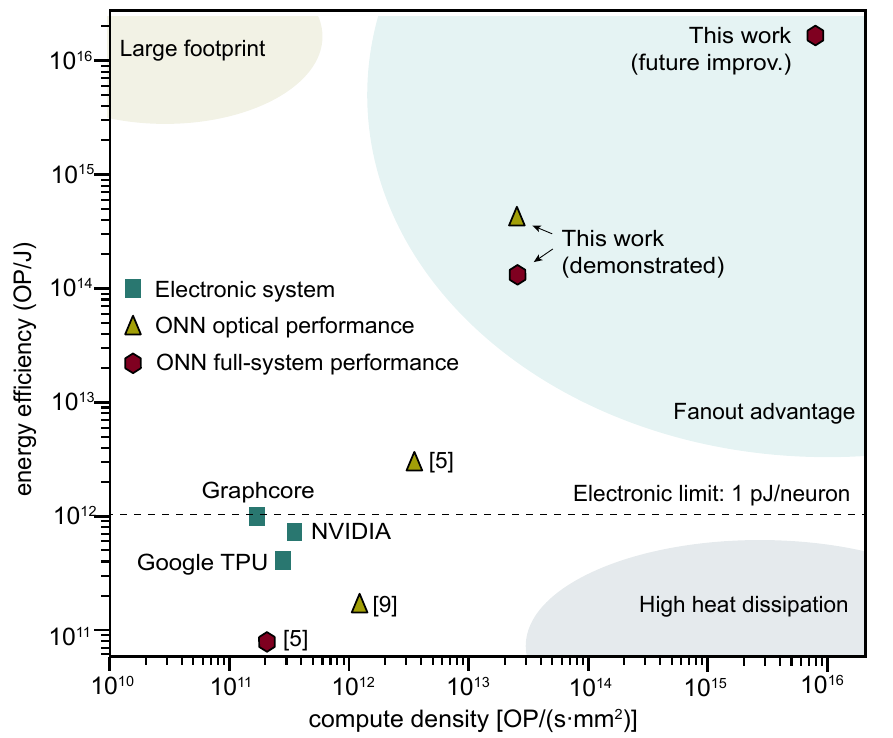}
    \caption{Comparison of state-of-the-art neural network hardware. The electronic systems Google TPU, NVDIA GPU and Graphcore are ASICs optimized for deep learning tasks with energy efficiency and compute density reaching 1 pJ/OP (Graphcore) and 0.35 TeraOP/(mm$^2\cdot$s) (NVDIA A100). For the ONN optical performance, the energy efficiency  accounts for the electrical power in laser generation and data encoding, while the compute density is calculated from the chip area for matrix operations. For ONN full-system performance, the energy consumption and compute density account for laser generation, data encoding, nonlinear activation, data readout, signal amplification, ADC, DAC and memory access. In this work, the energy bound due to electronics at $\sim$1 pJ/neuron is reduced with spatial fan-out and time-domain fan-in (Supplementary Table~I). The performance of each system is listed in Supplementary Table~II. }
    \label{fig:fig5}
\end{figure}

\noindent
\textbf{Energy efficiency.}
  Our system enabled efficient computing with low-energy VCSEL transmitters and optical parallelism. Supplementary Table I summarizes the energy consumption of each component. The clock rate of our system is 1 GS/s (as demonstrated in Fig.~\ref{fig:fig3}), limited by the VCSEL bandwidth. Due to the ultralow $V_{\pi}$=4 mV operation, data encoding with a VCSEL modulator consumes only 3.7 nanowatts (3.7 attojoules per symbol at 1 GS/s) (Methods), which is six orders of magnitude lower than that of previous ONN schemes with thermal phase shifters~\cite{SheHar17Deep}, microring resonators~\cite{Tait2017}, optical attenuators~\cite{Ashtiani2022} and EO modulators~\cite{Xu2021, Feldmann2021} that operate with several miliwatts of electrical power. As a result, the main optical energy consumption in our system is for laser generation, for which we note that VCSEL sources are efficient laser generators with wall-plug efficiency of 25~$\%$ in our demonstration and over 57~$\%$ in record~\cite{Wall_Plug_efficiency}. The theoretical lower bound to laser power is given by the number of photons required to produce a homodyne signal with sufficient bits of compute precision, which is ultimately limited by the required SNR from detection. Time integrating receivers (Methods), in contrast to conventional amplified detector, only read out after accumulating over $i$ time steps~\cite{sludds_netcast}, improving the SNR by a factor of $\sqrt{i}$. With off-the-shelf technology the thermal noise limit of computing from integration detection is 200 photons/OP (corresponding to 40 aJ/OP) as simulated in supplementary materials Sec. II. In our experimental demonstration, the VCSELs emit 100 $\upmu$W. The resulting optical energy efficiency, which includes the electrical power for laser generation and data modulation, is 2.5 fJ/OP (owing to the fan-out advantage) (methods). The performance of recent ONNs techniques has been summarized in Ref.~\cite{Ashtiani2022} and the results are plotted in Fig. \ref{fig:fig5}. Our optical energy efficiency of 2.5 fJ/OP is 140$\times$ and 1000$\times$ better than the state-of-the-art integrated ONN Ref.~\cite{Ashtiani2022} and Ref.~\cite{Feldmann2021}, respectively.
 
 Additional energy costs are from electronic digital-to-analog converters (DAC), analog-to-digital converters (ADC), signal amplification and memory access. The energy of DAC and memory access per use is reduced by a factor of $j$ due to spatial parallel processing with laser fanout. The read-out electronics, including ADCs, trans-impedance amplifiers and integrators, is triggered one once after time integration. Their energy cost per use is amortized by a total of 2$i$ operations in between. Thereby the full-system energy efficiency including both electronic and optical consumption is 7 fJ/OP (Supplementary Table~I), which is >100$\times$ better compared to the state-of-the art electronic microprocessors (Fig. \ref{fig:fig5}). Similar to the fanout of the input laser, the weight VCSELs can be spatially fanned out (with a factor of $k$) as demonstrated in Supplementary Sec~IV for batch operations, which reduces the energy for weighing to the same order of the input encoding.
 
\noindent
\textbf{Compute density.}
High compute density is achieved based on compact and dense VCSEL arrays in the three dimensional architecture. VCSELs are excellent candidate for high-density computing with a pitch of 80 $\mu$m per fabricated device, while nano/micro-pillar lasers with < 1 $\mu$m diameter \cite{nanolaser_connie} and <10 $\mu$m pitch \cite{Micropillar_VCSELs} have been demonstrated. The compute density in our system reaches $\rho$=25~TeraOP/(mm$^2\cdot$s) (Methods), which is about two orders of magnitude higher than that of electronic counter parts (Fig. \ref{fig:fig5}). Compared to electronic circuits, where improving throughput density is fundamentally challenging due to limited heat dissipation per chip area. Here with the improved energy efficiency in our VCSEL-ONN, higher throughput density is allowed. In other ONN configurations, high throughput density requires tiling photonic devices at high density, which often leads to severe crosstalk between neighbouring channels and decreased compute accuracy \cite{SheHar17Deep}. The channel crosstalk in our VCSEL-ONN is eliminated with VCSEL modulators of ultra-low $V_\pi$. 
  
\noindent
\textbf{Latency.}
Ultralow latency for nonlinear activation is achieved with incorporated detection-based nonlinearity. In our scheme, each detection event generates photon currents instantaneously, and the photon currents are accumulated in the time integrator for $i$ time steps before being read out. The transit time of photon electrons from the photo-diode to the charging capacitor, which leads to latency in standard photodetectors, is negligible compared to the integration time. So the latency due to nonlinear activation is negligible. The processing time is dominated by the data encoding and time integration, which could be as shown as 30 ns for a full-size MNIST image at the clock rate of R=25 GS/s (Fig.~\ref{fig:fig2}b). 

 \noindent

 \section*{Discussion}

  Harnessing the powerful scalability of the VCSEL platform, we demonstrated a homodyne-based ONN using >100 coherent VCSEL transmitters. Our benchmarking of digit classification achieves an accuracy of 93.1~$\%$ (over 98~$\%$ of ground truth). The full-system energy efficiency and compute density reaches 7 fJ/OP and 25 TeraOP/(mm$^2$$\cdot$s), both achieve 100$\times$ improvement compared to the digital hardware. In the near term, the weight matrix can be fanned out for matrix-matrix multiplication $X_{(k\times i)} W_{(i\times j)}=Y_{(k\times j)}$, where the overall throughput scales 2$\times$$j$$\times$$k$$\times$$R$, could exceed 50 PetaOP/s with present-day technology and practical hardware parameters ($j$=$k$=1000 and $R=25$~GS/s) \cite{miller}. Such a system is expected to operate multiply-accumulate with efficiency of $\sim$50 aJ/OP, limited by the memory access rather than optical energy consumption (supplementary Table I), and a compute density of 8 PetaOP/(mm$^2$$\cdot$s) is within reach due to the high clock rate and large fanout factor (Methods). Operating large neural network models (with $i$>10$^6$), the system allows optical energy efficiency of less than 1 photon/OP with a standard photo-detector at the room-temperature thermal-noise limit (Fig. S3), exhibiting ultrahigh optical energy efficiency. Moreover, based on the silicon platform, the VCSEL-ONN system can be integrated with electronic and photonic co-design using short wires to reduce electronic energy cost~\cite{packaging_wafer, Sun2015_wafer_packing}.
 
 Another crucial feature of our system is the rapid programmability with real-time updating weights, which is readily useful for neural network training. Weight streaming was a challenge to ONNs because it requires high-speed modulation for large-scale encoding of weights, leading to energy, speed and device overheads. Therefore almost all the ONNs operate with static weights \cite{Xu2021, wang2022optical, @LinScience2018}. It is made possible in our system due to the ultralow-energy consumption of VCSELs (<5 aJ/symbol). In addition, our technique of time-domain neuron encoding is scalable to tackle large models with up to billions of parameters (in 1 second compute time); further scaling on model size might be enabled with frequency multiplexing using VCSEL arrays of multiple wavelengths \cite{connie_VCSEL_multiple_wavelength}.
 
 In summary, our system has demonstrated promising potential to scale up computing power with many orders of magnitude improvement.  As the exponential scaling of neural network models has outpaced the development of electronic processors, this new type of optoelectronic processor may enable to continue the scaling of computing power in the post-Moore's law age.

\section*{Methods}
\label{Methods}

\noindent
\textbf{Device fabrication.} The VCSEL arrays are fabricated with a small pitch of 80~$\upmu$m to maximize the device density. The VCSEL cavities are based on semiconductor heterostructure microresonators with two AlGaAs/GaAs distributed Bragg reflectors (DBRs) as cavity mirrors and a stack of InGaAs quantum wells as gain medium~\cite{Heuser_2020}. The 5$\times$5 cavity arrays are patterned by UV lithography and etched by an inductively coupled plasma reactive ion beam. Each cavity with an outer diameter of 30~$\upmu$m is oxidized to an aperture of 4.5~$\upmu$m for suppressing the higher order transverse modes. To improve the laser stability, the whole chip is cladded with a polymer layer and the areas of the VCSEL cavities are reopened. The Au-deposited $p$-contact of each VCSEL is connected to a signal pad, which is wire bonded to a printed circuit board that links to external drivers. All the VCSELs share the common ground (golden bars in Fig. \ref{fig:fig2}d). The cross-section of the VCSELs are designed with 1$\%$-ellipticity, which allows polarized laser output with improved extinction ratio. 

\noindent
\textbf{Injection locking.} As shown in Fig. \ref{fig:fig2}f, the beam of the leader laser passing through a diffractive optical element (DOE), is reflected to the VCSEL arrays using a polarising beam splitter (PBS). At the Fourier plane of the coupling lens, the leader laser splits to a beam grid of spacing equal to the pitch of the VCSELs. The polarization of the PBS is aligned 45 degrees with respect to that of the VCSELs. So half of leader laser power is coupled to the VCSEL cavity, locking the phase of the remaining VCSEL oscillators. The other half, being reflected by the VCSEL front DBR, is rejected by the PBS to avoid falling onto the homodyne detectors leading to undesired interference. Such an injection-locking technique has been demonstrated recently to achieve high rejection ratio~\cite{Injection_lock_tech}. To simultaneously injection lock the whole array, we tune all the VCSELs to the wavelength of the leader laser by varying the bias voltage using a battery-supplied DC controller on each VCSEL channel. The injection lock is confirmed by monitoring the beatnote between the leader laser and each individual VCSEL. More details can be found in supplementary Sec.~VI. 

\noindent
\textbf{Phase modulation with injection locked VCSELs.} The phase ($\phi$) of an injection locked VCSEL is given by the frequency detuning between the leader laser and the VCSEL's free-running frequency, $\sin(\phi)\propto \delta_d/\delta_r$, where $\delta_d$ is the detuning and $\delta_r$ is the injection-locking range. Consistent with the theory of injection locking, the range $\delta_r$ is proportional to the square root of the injecting power (supplementary Fig. S8), so a small V$_\pi$ (on the mV range) is achieved by reducing the injecting power (to about 1~$\upmu$W per VCSEL). The result is similar to that in Ref.~\cite{VCSEL_modulator_Hoghooghi:10}. The frequency response of the injection-locked VCSELs in the thermal region (<10 MHz) is about 10 dB stronger than that in the free-carrier region (Fig. S8d). To decouple from the thermal effect, we modulate the data with a high-frequency local oscillator and demodulate it from the homodyne signal (supplementary Fig. S10). This data modulation scheme is not needed with VCSELs operating at higher data rates, as experimentally demonstrated in ref.~\cite{VCSEL_OAWG_Bhooplapur}.

\noindent
\textbf{Time integrating receiver.} The photon receiver is connected to a homemade switched integrator charge amplifier based on the integrated circuit IVC102 from Texas Instruments. Its performance in ONNs has been discussed thoroughly in our previous work \cite{sludds_netcast}. The capacitor in the integrator accumulates charges when the switch is on and output an integrated voltage $V_{int}\propto \sum_i f(X_i^n, W_{ij}^n)$ when switched off.

\noindent
\textbf{Training}. The training model is custom-designed to implement tailored nonlinearity. It consists of one input layer, two fully connected hidden layers, and a output layer (Fig. \ref{fig:fig4}). The input layer consists of 784 neurons, corresponding to a full-size MNIST image with a handwritten digit. Two fully connected hidden layers are used. In each layer, the matrix-vector multiplication is computed with our custom nonlinear synaptic weighting function $f_{NL}(W_j^n, X^n)$. The output layer consists of 10 neurons, each neuron represents a digit (from 0 to 9); the prediction result of which digit is given by the number of neuron which has the largest value. We implement batch normalization~\cite{batch_normalization} in each layer to accelerate the training process. The initial weights~\cite{weight_init} are optimized such that the integrated partial sums in each layer converge to a mean value around 0, which allows one to maximize the experimental signal level without being bounded by the dynamic range limit. We utilize the cross-entropy loss function and retrieve the gradients in each iteration in a model implemented in PyTorch~\cite{Pytorch}. A large learning rate is set to start the training and is gradually reduced to optimize the accuracy.  

\noindent
\textbf{System performance} The VCSEL emits $P_i=100$ $\upmu$W optical power, while consuming $P_b$=400 $\upmu$W electrical power. The injection-lock power is $P_\textsc{inj}=1~\upmu$W per VCSEL. The power for data modulation is $P_\textsc{m}=V_\pi^2/R_\textsc{VCSEL}$=3.6 nW, with $R_\textsc{VCSEL}$=4.3 k$\Omega$ and V$_\pi$=4 mV. The maximum clock rate of the system is $R$=1 GS/s. With a fanout factor $j=9\times$9, the optical energy efficiency is ($P_b+P_\textsc{inj}+P_m)/(2jR)$=2.5 fJ/OP, dominated by the laser power. The full-system energy efficiency is 7 fJ/OP, as discussed in Supplementary Table~II. For compute density, the chip area is dominated by the VCSEL transmitters while silicon detector arrays are compact (e.g., <2$\times$2 $\upmu$m$^2$ pixel size for an image sensor in cellphones) and the phase mask can be imprinted directly on the VCSEL output facet. The area of each VCSEL device is $a=80\times 80~\upmu\text{m}^2$. The compute density with $\rho$=2$jR$/($a$)= 25~TeraOP/(mm$^2\cdot$s), accounted for the input laser which is fanned out. In future development, when the weighting VCSELs can be fanned out by a factor of $k$ (e.g., $k=j=9\times 9$ in supplementary Fig.~S5) for batch operation, the energy consumption and compute density of weighting is the same as the input. For future improvement with $R$=25 GS/s and $j$=1000, the compute density is expected to reach $\rho$=2$jR$/($a$)= 7.8~PeraOP/(mm$^2\cdot$s).

\section*{Acknowledgements}
\label{sec:acknowledgements}

This work is supported by the Army Research Office under grant number W911NF17-1-0527, and NTT Research under project number 6942193 and the NTT Netcast award 6945207. We also acknowledge financial support for the Volkswagen Foundation via the project NeuroQNet 2. We thank Lamia Ateshian for assembling an initial setup for testing VCSEL samples. The authors would like to thank Prof. David A.\ B.\ Miller of Stanford University for the informative discussions on high-speed and low-energy interconnects. We thank Dr. Nicholas Harris and Dr.\ Darius Bunandar of Lightmatter for the discussions on the scalibilities of optical neural networks. D.E.\ would like to thank Prof.\ Seth Lloyd of MIT for discussions. 

\section*{Author Contributions}
\label{sec:author_contributions}
  Z.C., R.H., D.E.\ conceived the experiments. Z.C.\ performed the experiment, assisted by A.S., R.D., I.C..\ A.S. conducted high-speed measurements on the VCSEL transmitters. R.D.\ created the software model for neural network training. I.C.\ performed electronic packaging on VCSEL arrays. A.S.\ and L.B.\ assisted with discussions on the experimental data. T.H., N.H., J.L., and S.R.\ designed and fabricated the VCSEL arrays and characterized their performance. R.H.\ and D.E.\ provided critical insights on experimental implementation and result analysis. Z.C.\ wrote the manuscript with contributions from all the authors.

\section*{Competing Interests}
\label{sec:competing_interests}
Z.C, D.E and R.H have filed a patent related to VCSEL ONNs, bearing application No.63/341,601. D.E. serves as scientific advisor to and holds equity in Lightmatter Inc. Other authors declare no competing interests.

\section*{Data Availability}
\label{sec:data_availability}
The data supporting the claims in this paper is available upon reasonable request.

\section*{Correspondence}
\label{sec:correspondence}
Requests for information should be directed to Zaijun Chen (zaijunch@usc.edu).

\bibliography{bibliography.bib}{}

\end{document}


\title{Supplementary Materials: Deep Learning with Coherent VCSEL Neural Networks}
\author{Zaijun Chen$^{1}$}
\author{Alexander Sludds$^{1}$}
\author{Ronald Davis$^{1}$}
\author{Ian Christen$^{1}$}
\author{Liane Bernstein$^{1}$}
\author{Tobias Heuser$^{2}$}
\author{Niels Heermeier$^{2}$}
\author{James Lott$^{2}$}
\author{Stephan Reitzenstein$^{2}$}
\author{Ryan Hamerly$^{1,3}$}
\author{Dirk Englund$^{1}$}

\affiliation{$^{1}$Research Laboratory of Electronics, MIT, Cambridge, MA 02139, USA}
\affiliation{$^{2}$Fakultät II Institut für Festkörperphysik Sekretariat, Technische Universität Berlin, Berlin, 10623, Germany}
\affiliation{$^{3}$NTT Research Inc., PHI Laboratories, 940 Stewart Drive, Sunnyvale, CA 94085, USA}



\maketitle
\tableofcontents

\newpage

\section{Homodyne detection} \label{sec:homodyne_detection}
 
\begin{figure*}[ht]
    \centering
  \includegraphics[width=140mm]{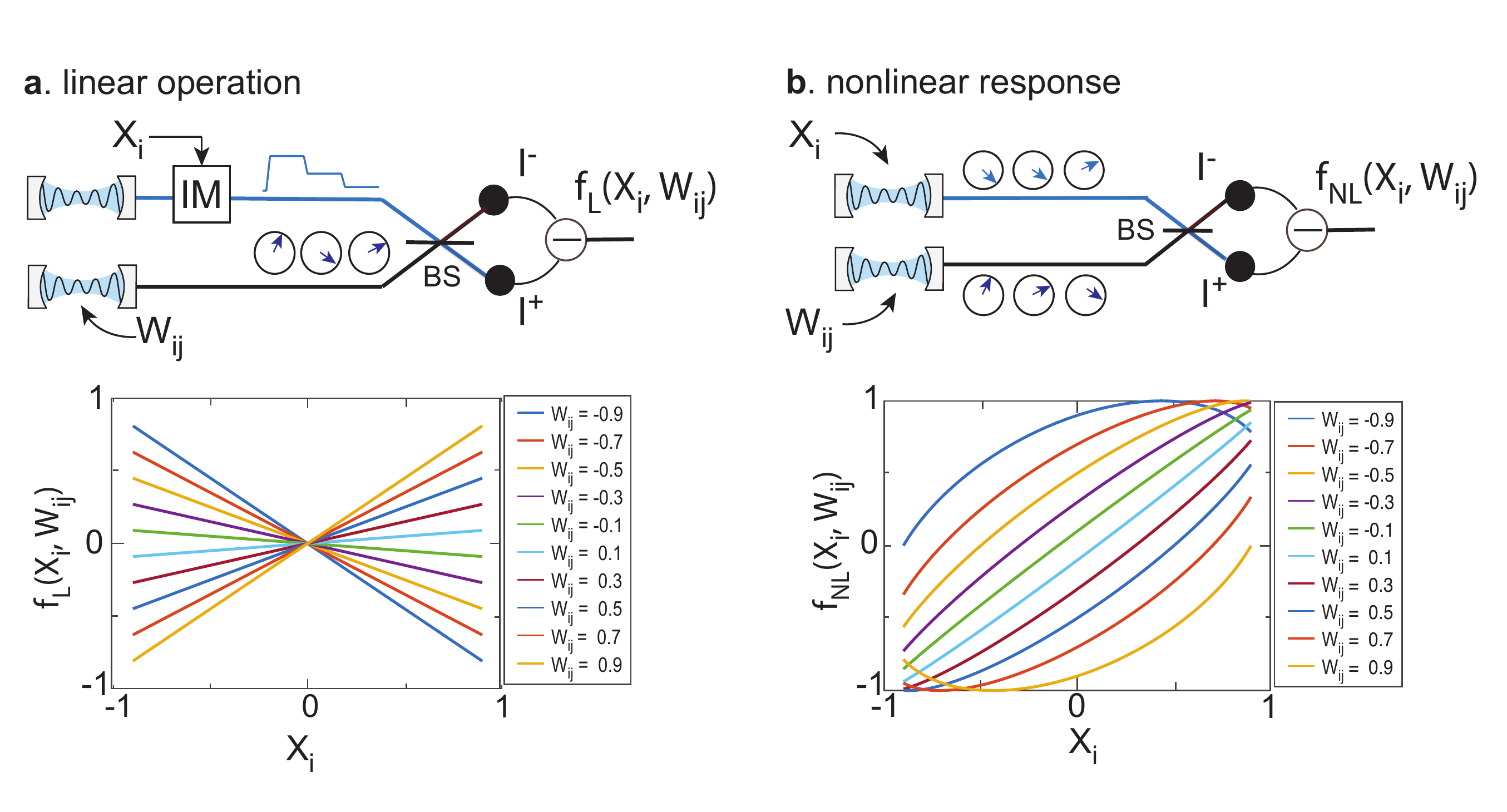}
    \caption{Linear and nonlinear operation based on homodyne coherent detection. \textbf{a.} Linear input-output response when the input data is amplitude encoded. \textbf{b.} Nonlinear activation when the input data is phase encoded. The input-output nonlinearity that increases at high weights is programmable by setting the phase of the weight laser. }
    \label{fig:linear_and_nonlinear}
 
 \end{figure*}

 \noindent
 Matrix operations in our VCSEL-ONN are based on homodyne detection. A compute unit for vector-vector multiplication composed of two VCSEL emitters is shown in Fig.~S\ref{fig:linear_and_nonlinear}. We encode the input vector $X^n$ (of size 1$\times$$i$) and a weight matrix to $W^n$ (of size $i\times$$j$) in $i$ time steps to the amplitude or phase of the VCSEL emission. The two lasers are coherent by means of injection-locking using a leader laser emitting at frequency $\omega$. The electric field of the two laser oscillators is\\
 \vspace{-5mm}
 \begin{equation}\label{eq:laser_fields_X}
 E_X(t)=A_Xe^{-i\omega t+\phi_X},
 \end{equation}
  \vspace{-14mm}
 \begin{equation}\label{eq:laser_fields_W}
 E_W(t)=A_We^{-i\omega t+ \phi_W},
 \end{equation}

where $A_X$ and $\phi_X$, $A_W$ and $\phi_W$, respectively, are the amplitude and phase of the input laser and the weight laser. The beams of the two VCSELs are overlapped using the beamsplitter, which induces a phase delay of $\pi/2$ to the reflected beam. The homodyne receiver detects photoncurrents as

 \vspace{-7mm}
 \begin{equation}\label{eq:I+}
 I^+\propto\ [E_We^{-i\pi/2}+E_X]\times [E_We^{-i\pi/2}+E_X]^*, 
 \end{equation}
  \vspace{-14mm}
 \begin{equation}\label{eq:Ia}
 I^-\propto\ [E_W+E_Xe^{-i\pi/2}]\times [E_W+E_Xe^{-i\pi/2}]^*, 
 \end{equation}
 
 Plugging Eq. 1 and Eq. 2 into Eq. 3 and Eq. 4, one obtains
 
  \vspace{-7mm}
 \begin{equation}\label{eq:I-}
 I^+\propto\ \lvert A_X\rvert^2 + \lvert A_W\rvert^2 + 2A_XA_W\sin[\phi_W-\phi_X], 
 \end{equation}
   \vspace{-14mm}
  \begin{equation}\label{eq:Ib}
 I^-\propto\ \lvert A_X\rvert^2 + \lvert A_W\rvert^2 - 2A_XA_W\sin[\phi_W-\phi_X], 
 \end{equation}
 
 When balance detection is used, it reads the differential currents,
 \vspace{-2mm}
 \begin{equation}\label{eq:balance_detection}
  \Delta I(t)=I^+(t)-I^-(t)\propto A_XA_W\sin[\phi_W-\phi_X]
\end{equation}

where the non-interference terms in Eq. 5 and 6 are cancelled and the interference term is enhanced by a factor of 2, due to the opposite phase induced by the beam-splitter.

Depending on the data modulation scheme of the input laser, the system allows linear and nonlinear operations. In both cases, the weight lasers are phase modulated following the format, $\sin[\phi_W(t)]\propto W_{ij}$.
 
 \textbf{Linear operation} is activated when the input vector is amplitude encoded with $A_X(t)=X_i$, as shown in Fig.~S\ref{fig:linear_and_nonlinear}a. Because the VCSELs for encoding of weights are phase-only modulators, $A_W$ is constant. When the phase of the input laser is set $\phi_X=0$ (two VCSEls are in phase), the interference signal is simplified to
 
 \vspace{-7mm}
 \begin{equation}\label{eq:diff}
 \Delta I(t)\propto A_X(t)\sin[\phi_W(t)]=X_i W_{ij},
 \end{equation}
 
where the multiplication of the input value $X_i$ and a weight value $W_{ij}$ is achieved in the homodyne product.  
 
 \textbf{Nonlinear operation} is allowed (Fig.~S\ref{fig:linear_and_nonlinear}b) when the input vector is phase encoded, $\sin[\phi_X(t)]\propto X_{i}$. Here both laser fields are with phase-only modulation, so the non-interference terms are direct currents that can be coupled out from the AC term. The inference can be detected with a balanced detector for signal-to-noise ratio (SNR) improvement, or with a single detector for system simplicity. The generated photocurrent on the detector is
 
  \vspace{-7mm}
 \begin{equation}\label{eq:balanced-currents}
 \Delta I(t)\propto I(t)^+\propto \sin[\phi_W(t)-\phi_X(t)]
 =W_{ij}\sqrt{1-X_i^2}-X_{i}\sqrt{1-W_{ij}^2},
 \end{equation}
 
 where $\sin[\phi_W(t)-\phi_X(t)]=\sin[\phi_W(t)]\cos[\phi_X(t)]-\cos[\phi_W(t)]\sin[\phi_X(t)]$, $\sin^2(\phi)+\cos^2(\phi)=1$, $\sin[\phi_X(t)]\propto X_{i}$ and $\sin[\phi_W(t)]\propto W_{ij}$ are used. The input-output response of the linear and nonlinear models is simulated in Fig. S\ref{fig:linear_and_nonlinear}.

\newpage

\section{ Energy consumption} \label{Energy_consumption}

A fundamental limit to the optical energy consumption is given by the optical power required to achieve a desired signal-to-noise ration (SNR) in homodyne detection, which sets the compute bits of precision. In this section, we model the signal and noise sources in the system and validate the model with our experimental results. Then we apply the model to discuss the lower bound of optical energy consumption for our current system limit and future improvement. 

\noindent
\textbf{1. Signal-to-noise analysis}

\begin{figure*}[h!]
    \centering
    \includegraphics[width=180mm]{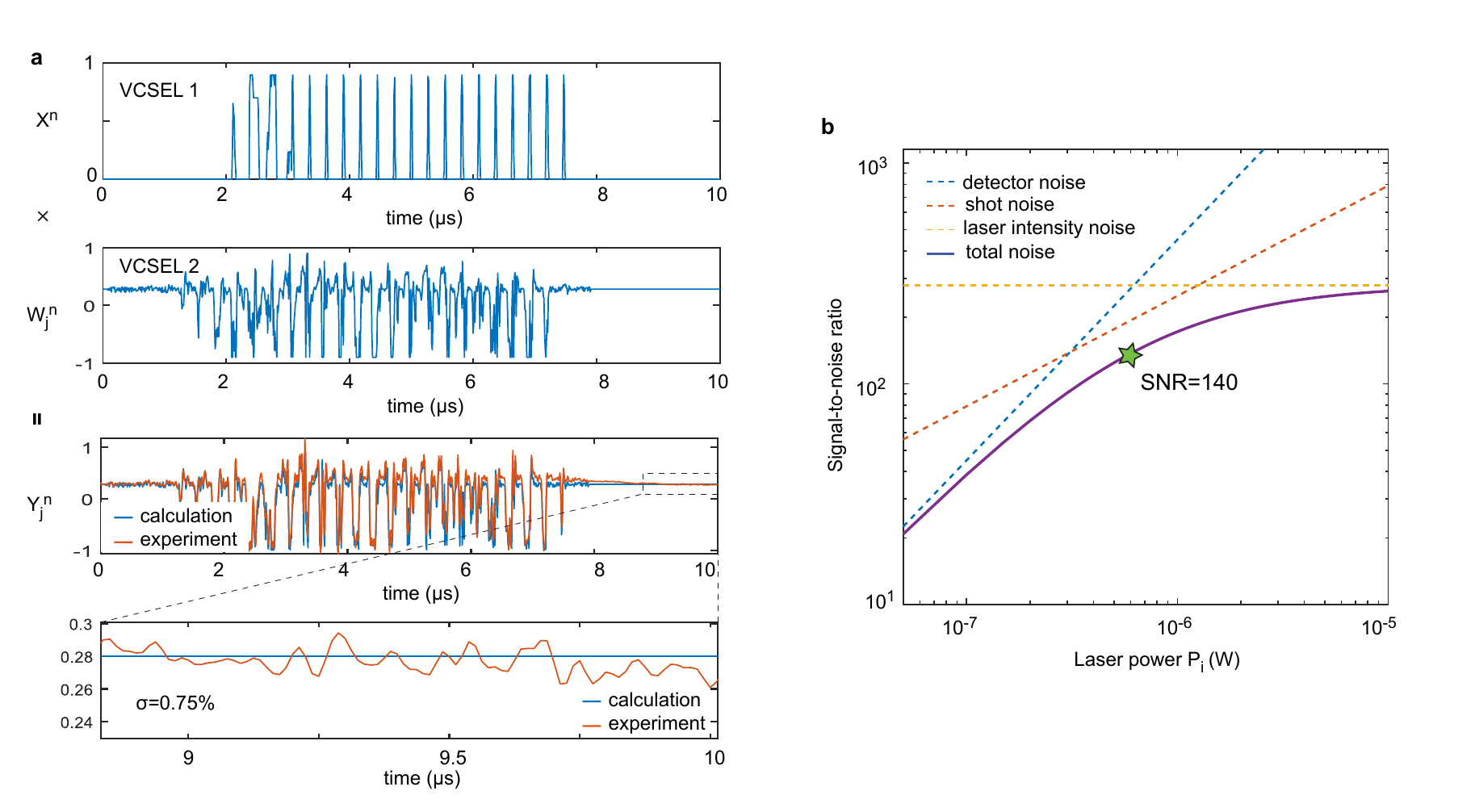}
    \caption{SNR analysis. \textbf{a.} Experimental demonstration of homodyne interference. Multiplying a image data (encoded with VCSEL 1) with a weight vector (Encoded in VCSEL 2). The homodyne signal agrees well with the calculated result. The SNR in the homodyne signal is given by the standard deviation of the baseline normalized to the peak-to-peak amplitude 1/$\sigma$=135.  \textbf{b.} the SNR is modeled with the experimental conditions: the NEP of the detector is 5 pW/$\sqrt{Hz}$, and the laser relative intensity noise is -145 dBc/Hz, b=2 for unbalanced detector, $\gamma$=$j$=$9\times$9, $\nu$=307.5 THz, and $\eta=0.65$.  The homodyne detector receives $P_W$=50 $\mu$W and $P_X$=0.6 $\mu$ W (50 $\mu W$ fans out to $\gamma$=$j$=9$\times$9 copies). The data clock rate is $R$=100 MS/s, corresponding to $t_c$=1/$R$=10 ns.}
    \label{fig:signal_to-noise}
\end{figure*}

The SNR of optical interference with two laser fields has been well understood in Ref. \cite{Newbury:10} and is adapted here.  We denote $P_X$ and $P_W$ as the power of the input laser and the weight laser on the homodyne receiver. The root mean square amplitude of the interference signal is $S=\sqrt{2}\eta \sqrt{P_XP_W}=\sqrt{2\gamma}\eta P_i$, where $\eta$ is the quantum efficiency of the photo-detector. For simplicity, we denote $\gamma=P_W/P_X$ and $P_X=P_i$. We include the detector thermal noise, the photon shot noise and the laser intensity noise in the model. The total noise amplitude is  

\begin{equation}\label{eq:SNR}
N_t=\sqrt{[(\eta\textsc{NEP})^2+2(1+\gamma)h\nu \eta P_i+b(1+\gamma^2)(\eta P_i)^2\textsc{(RIN)}]B}
\end{equation}

where NEP is the noise equivalent power of the photo-receiver. $h$ is the Planck's constant, $\nu$ is the frequency, RIN is the relative laser intensity noise. b=1 accounts for the reduction of relative intensity noise due to the balance detection and 2 for unbalanced detection. The noise spectrum is integrated over the effective detection bandwidth B=1/(2T), with T being the acquisition time. The average uncertainty in our homodyne signal is,

\begin{equation}\label{eq:homodyne_SNR}
\sigma_H=N/S_t=\frac{1}{2\sqrt{T}}\bigg[\frac{\textsc{NEP}^2}{\gamma P_i^2}+\frac{4c_\gamma h\nu}{\eta P_i} +2bc_{\gamma^2}\textsc{(RIN)}\bigg]^{-1/2}
\end{equation}

where $c_\gamma=(1+\gamma)/(2\gamma)$ and $c_{\gamma^2}=(1+\gamma^2)/(2\gamma)$ accounts for the power ratio of the two lasers. The SNR can be calculated as $1/\sigma_H$.

In the implementation of neural network model (Fig.~S\ref{fig:signal_to-noise}), the data is read out in every time step T=$t_c$=1/R=10 ns. The SNR in the experimental homodyne signal is 135, which agrees well with the theoretical SNR of 140 predicted by the model (Fig.~S\ref{fig:signal_to-noise}). The main contribution of noise source is shot noise.

\noindent
\textbf{2. Analysis of system performance}

\begin{figure*}[ht]
    \centering
    \includegraphics[width=160mm]{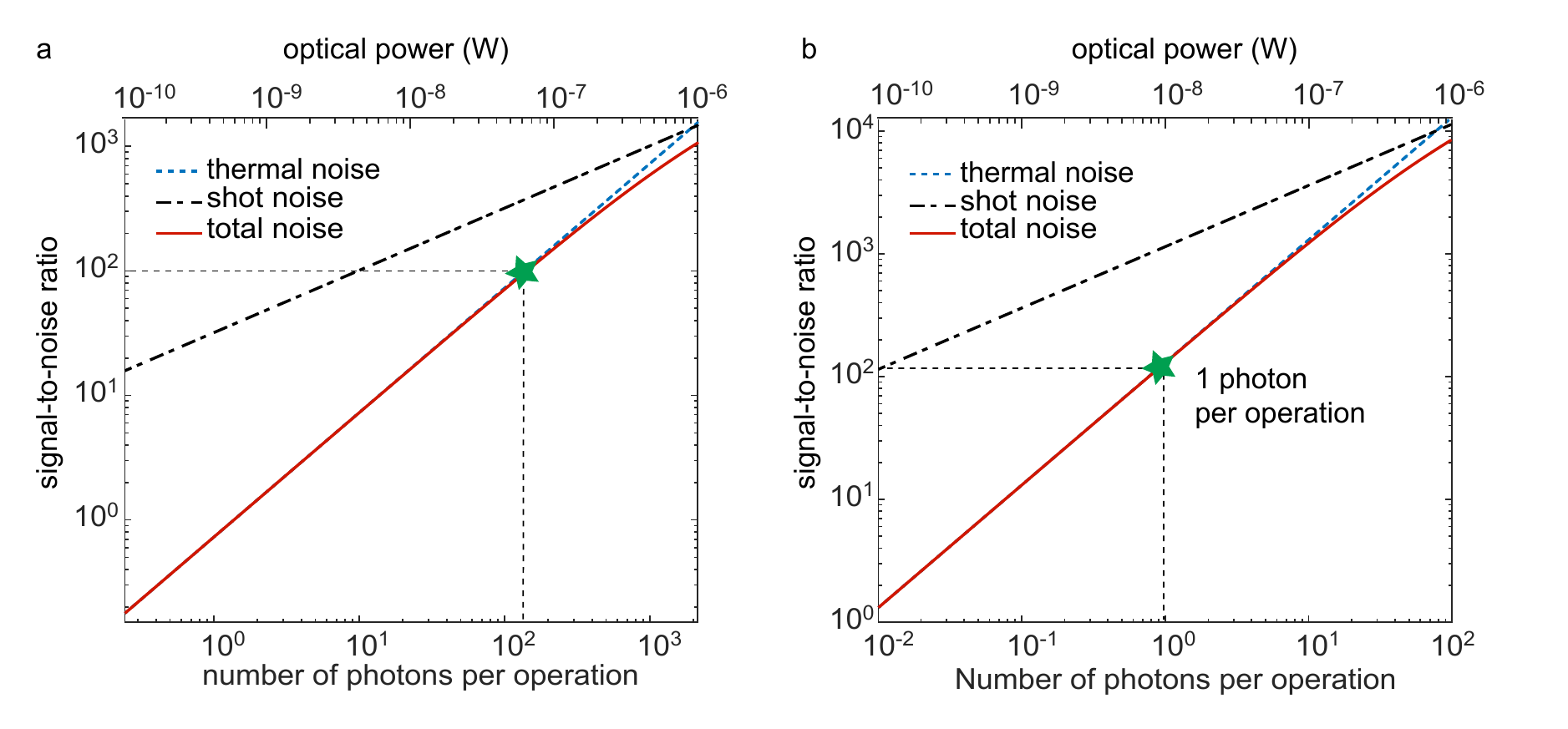}
    \caption{SNR as a function of input laser power (or number of photons per operation). The SNR is calculated over $i$ integrating time steps. \textbf{a} Experimental conditions. The clock rate is $R$=1 GS/s and the signal is accumulated over $i$=784 time steps. The detector NEP is 1 pW/$\sqrt{\textsc{Hz}}$.  b=2 for unbalanced detector, $\gamma$=1, $\nu$=307.5 THz, and $\eta=0.65$. \textbf{b.} Performance with high-speed VCSELs. With $R$=25 GS/s and $K$=$10^6$, the acquisition time is T=400 ns per readout. Less than 1 photon per operation allows for a compute precision of about 6$\sim$7 bits. }
    \label{fig:SNR_photon_number}
\end{figure*}

We analyze the system performance utilizing the model above. Different from the readout of every time steps in previous section (Fig.~S\ref{fig:signal_to-noise}), here we use an integrating receiver, which only reads out the integrated value after $i$ time steps. The data acquisition period is $T=i\cdot$$t_c$.  The factor of $i$ longer integration time results in a factor of $\sqrt{i}$ improvement in the SNR, as we use amplitude representation. Eq 11 is modified to

\begin{equation}\label{eq:integration}
\sigma_I=N/S=\frac{1}{2\sqrt{i\cdot t_c}}\bigg[\frac{\textsc{NEP}^2}{\gamma P_i^2}+\frac{4c_\gamma h\nu}{\eta P_i} +2bc_{\gamma^2}\textsc{(RIN)}\bigg]^{-1/2}
\end{equation}

Fig.~S\ref{fig:SNR_photon_number} shows the evolution of SNR with optical power. In \textbf{a}, we analyze the theoretical energy limit with our experimental conditions. The contribution of laser intensity noise in the plotted power range is negligible compared to the detector noise and shot noise. With time integration, the detector bandwidth is reduced by a factor of $i$ (e.g., B=1 MHz is sufficient for $R$=1 GHz and $i$=1,000). The thermal noise on detector is reduced accordingly with $\sqrt{i}$. We use NEP=1 pW/$\sqrt{\textsc{Hz}}$ in the calculation due to the reduced bandwidth.  As our system is designed to operate at room temperature, the SNR at low power level is dominated by thermal noise, which is different from the results from Ref. \cite{wang2022optical_1, sludds_netcast_1}, where the detector is cooled (and that consumes energy) to operate at the shot noise limit. As shown in Fig.~S\ref{fig:SNR_photon_number}a, a SNR of 100 is obtained with 200 photons per operation, corresponding to the energy efficiency $\epsilon$=40 aJ/OP. The SNR of 100 corresponds to about 7 bits of precision, which is sufficient for most neural network tasks \cite{quantized_low_precision}. In the future, with high-speed VCSELs that operates at $R$=25 GS/s and with larger model sizes that integrate over a million neurons $i$=10$^6$, less than 1 photon per operation ($P_i$=10 nW) is achievable (Fig.~S\ref{fig:SNR_photon_number}b), corresponding to 0.2 aJ/OP.

\noindent
\textbf{3. system power consumption}

We analyze the energy consumption of each component in the system, including both the optical energy and the consumption of electronic components, as shown in Table \ref{tab:Energy_system}. 

\vspace{1cm}
\begin{table}[ht]
\centering
\caption{Energy budget for our experimental apparatus and future improvement using conventional technology.}

    \begin{tabular}{|p{5cm}|p{2.6cm}|p{1.6cm}|p{2cm}|p{2.6cm}|p{1.6cm}|p{2cm} | }
    \hline
    \multirow{2}{*}{\centering Parameter}
     & \multicolumn{3}{|c|}{system performance} & \multicolumn{3}{|c|}{future improvement} \\
     \cline{2-7}
                                                                & \centering Value                           & \centering $^1$Fan-out       & \centering $^2$Energy/OP & \centering Value & \centering Fan-out &  Energy/OP  \\
    \hline
     VCSEL bias voltage ($V_b$)                        & \centering 1.3 V                         &  &  &  &  &\\
     VCSEL bias current ($I_b$)                        & \centering 300 $\mu A$                         &  &  &  &  &\\
     Laser power ($P_i$)                          & \centering 100 $\mu$W                      &                            &                    & \centering 10 $\mu W$  &  & \\
     Laser wall-plug efficiency$^3$ ($\xi$)           & \centering 25~$\%$                         &  &  & \centering 25~$\%$ &  &\\
     $\pi$ phase shift voltage ($V_{\pi}$)        & \centering 4 mV                         &  &   & \centering 4 mV &  &\\
     Datarate ($R$)                               & \centering 1 GS/s                          &                           &                      & \centering 25 GS/s  &  & \\
     Injection power ($P_{inj}$)      & \centering 1 $\mu$W           &       &       & \centering 1 $\mu W$  &  & \\
     Power EO modulation$^4$ ($P_m$)           & \centering 3.7 nW   &   &                    &  &  & \\
     Power for optical energy$^5$ ($E_{opt}$)             & \centering  400 fJ/symbol & \centering $j$=9$\times$9   
     & \centering 2.5 fJ &\centering 1.6 fJ/symbol & \centering $j$=32$\times$32 &  0.8 aJ \\
     Nonlinear activation$^6$ ($E_{\textsc{NL}}$)     & \centering (400 fJ/symbol)                       & \centering ($j$=9$\times$9)   & \centering (2.5 fJ) & \centering 1.6 fJ/symbol   &  \centering $j$=32$\times$32  &(0.8 aJ)\\
     ADC$^7$  ($E_{\textsc{ADC}}$)                    & \centering $\sim$ 1~pJ/use \cite{MurmannDAC}   & \centering $i$=28$\times$28 & \centering 0.5 fJ & \centering $\sim$ 1~pJ/use   &\centering $i$=$10^6$ & 0.5 aJ\\
     Integrator  ($E_{\textsc{INT}}$)             & \centering $\sim$ 1~fJ/use \cite{Yang2019}     & \centering $i$=28$\times$28 & \centering 1 aJ & \centering $\sim$ 1~fJ/use & \centering $i$=$10^6$ & 0.5 zJ \\
     Energy of TIA$^7$ ($E_{\textsc{TIA}}$)           & \centering $\sim$ 1~pJ/use \cite{TIA}          & \centering $i$=28$\times$28 & \centering 0.5 fJ &  \centering $\sim$ 1~pJ/use & \centering $i$=$10^6$ & 0.5 aJ \\
     DAC$^8$  ($E_{\textsc{DAC}}$)                & \centering $\sim$ 0.5~pJ/use \cite{DAC_5GHZ}    & \centering $j$=$9\times$9 &     \centering 3 fJ & \centering $\sim$ 1.6~aJ/use   &\centering $j$=32$\times$32 & 0.8 zJ\\
     Memory access$^9$  ($E_{\textsc{MEM}}$)      & \centering $\sim$ 100 fJ/access   & \centering $j$=$9\times$9 &     \centering 0.6 fJ & \centering $\sim$ 100~fJ/access   &\centering $j$=32$\times$32 & 50 aJ\\
    \hline
     Total energy ($E_{tot}$)           &          &  &\centering $\sim$7 fJ  &  &  & $\sim$50 aJ \\
    \hline
    \end{tabular}
\label{tab:Energy_system}
\end{table}

\noindent
1. Fanout: $j$=9$\times$9 and $j=32\times$32 are spatial fanout with phase mask. Data encoding to the laser field consumes energy for memory access and digital-to-analog converter (DAC). This energy cost is split to $j$ times when the beam is fanned out for parallel operations. Similarly, the analog-to-digital converter (ADC), trans-impedance amplifier (TIA) and integrator are triggered once after integrating $i$ time steps, so their operation rate is factor of $i$ slower than the clock rate, and their energy consumption per use is divide by the $2\cdot$$i$ operations. $i$=28$\times$28 corresponds to time integration over a image of 28$\times$28 pixels.

\noindent
2. OP: operation in multiply and accumulate (MAC).  Every symbol in data encoding, and every use of ADC, TIA, integrator, DAC, memory access computes 1 MAC=2 operations (OP).

\noindent
3. The wall plug efficiency is $\xi=P_i/P_b=25~\%$. The electrical power for laser generation is determined by product of the bias voltage and the current flowing through the VCSEL $P_b$=$V_b\times$$I_b$=390 $\mu$W.

\noindent
4. The power used for electro-optic (EO) data modulation is $P_m$=$V_\pi^2$/$R_\textsc{VCSEL}$=3.7 nW with the VCSEL resistance $R_\textsc{VCSEL}$=$V_b$/$I_b$=4.3 k$\Omega$. The corresponding energy per operation is $E_m$=$P_m$/$R$=3.7 aJ/sample with $R=$1 GS/s.

\noindent
5. This is the total electrical power of laser generation, injection locking and data modulation, $E_{opt}=(P_b+P_{\textsc{inj}}/\xi$+$P_m$)/(2$R$). Here the leader laser for injection locking is a VCSEL with the same wall-plug efficiency $\xi$.  The power for injection lcoking and data modulation are negligible compared to that for laser generation. With future improvement with $R$=25 GS/s,  $i$=10$^6$ and $j$=32$\times$32, the laser consumes 40 $\mu$W electrical power and emits 10 $\mu$W optical power, which fans out to 10 nW ($j$=32$\times$32) per homodyne detector. This power is sufficient for a SNR=100 in the VCSEL-ONN (see previous section), the optical power consumption reaches $\epsilon_p$=0.8 aJ/OP. 

\noindent
6. The homodyne nonlinearity consumes the same power as the optical energy.  

\noindent
7. Standard CMOS ADC and TIA operates at 1 pJ/bit, corresponding to 0.5 pJ/OP. Our system performs signal amplification and ADC after integrating over $i$ time steps. So the ADC and TIA energy costs are reduced by a factor of $i$. 

\begin{figure*}[ht]
    \centering
    \includegraphics[width=100mm]{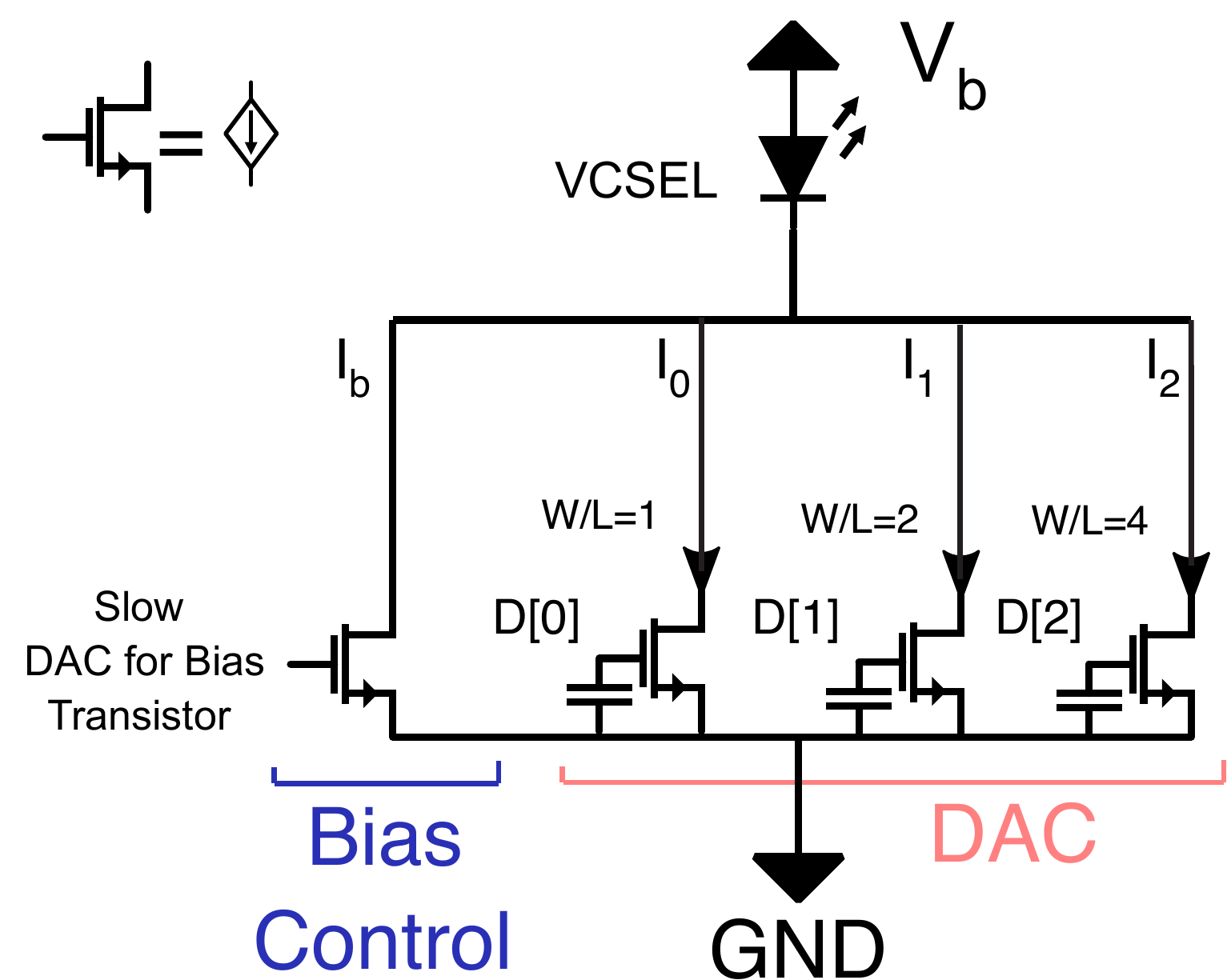}
    \caption{Proposed CMOS circuit for driving VCSELs at ultralow $V_{\pi}$. The circuit consists of a bias controller for laser generation and a bias transistor for wavelength tuning in injection locking. The VCSEL is forward biased with a voltage of $V_b$ and the current $I_b$ flows to a large transistor to tune the VCSEL wavelength for injection locking. As the current of the transistor $I_{transistor}\propto W_T/L_T$, where $W_T$ is the width and $L_T$ is the length, the size of the DAC transistors is designed to be hundreds of times smaller than that of the bias transistor. The small DAC transistors pull currents from $I_b$, producing small-signal modulation for low $V_{\pi}$ modulations.}
    \label{fig:DAC}
\end{figure*}

\noindent
8. Conventional CMOS digital-to-analog converters (DAC) operates at 0.5 pJ/use \cite{DAC_5GHZ}, corresponding to 0.25 pJ/OP. Each DAC generates a neuron activation that is fanned out to 9$\times$9 copies, leading to 3 fJ/OP after fanout. This posts a limitation to the existing system. However, these CMOS DACs are designed to output at voltage >1 V \cite{DAC_5GHZ}, which is practically not required in our system because our VCSELs operate with $V_{\pi}$=4 mV. As $E_{DAC}$ scales quadratically with operation voltage \cite{DAC_understanding}, it can significantly reduced at low voltage operation. However, modern DACs are not designed to operate with such a low supply voltage. We propose a current-drive DAC circuit as Fig. S\ref{fig:DAC}, where the DAC supplies the power required for EO modulation and consumes only $E_\textsc{DAC}=CV^2/2$=1.6 aJ/use, where $C$=200 fF/bit is the capacity for 1-mm wire, and $V$=0.4 mV is the $V_{\pi}$. This DAC uses a larger bias-control transistor to bias the VCSEL into lasing. Smaller transistors, sized such that the current they pull is staggered in powers of 2, convert a digital signal to a current which has the same effect as a 4 mV peak to peak drive voltage.

\noindent
9. Another energy cost is for delivering the signal from memory to the DAC. This charges the electronic wires, with $E_\textsc{MEM}=CV_\textsc{MEM}^2/2$=100 fJ/access, where $C$=200 fF/bit is the capacity for 1-mm wire, and $V_\textsc{MEM}$=1 V is the voltage for switching gates of transistors in electronic memory.

\section{comparison of state-of-the-art computing hardware}.    \label{sec:comparison_of_stete_of-the-art}

The performance of digital computers and ONN systems is based on the survey of Ref. \cite{Survey_hardware} and the analysis in Ref. \cite{Ashtiani2022}, respectively. 

\vspace{1cm}
\begin{table}[ht]
\centering
\caption{Performance of state-of-the-art computing hardware.}

    \begin{tabular}{|p{2cm}|p{5.2cm}|p{3cm}|p{2.6cm}|p{4cm}|}
    \hline
    \multicolumn{2}{|c|}{Hardware} & compute density (TeraOP/(mm$^2\cdot$s)) & Energy efficiency (TeraOP/J) & Comments \\
    \hline
    \multirow{3}{4em}{Digital computer} &Google TPU~\cite{TPU_ASIC} & 0.28 & 0.4  &\\
    &NVIDIA A100~\cite{A100, Survey_hardware}  & 0.35 & 0.72  & \\
    &Graphcore IPU2~\cite{GraphCore, Survey_hardware} & 0.17 &1.0 &\\
    \hline
    \multirow{4}{4em}{Optical Neural network} & Photonic tensor core~\cite{Feldmann2019} & 1.2 & 0.4 & optical energy$^d$  \\
    & Photonic deep neural network~\cite{Ashtiani2022}  & 3.5 & 2.9$^a$  & optical performance  \\
    & Photonic deep neural network~\cite{Ashtiani2022}  & 0.03$^b$ & 0.07$^c$  & full-system performance$^e$  \\
    & This work (now) & 25 & 140 & full-system performance   \\
    & This work (now) & 25 & 400 & optical performance   \\
    & This work (future improvement) & 8000 & 20000 & full-system performance \\
    \hline
    \end{tabular} 
\label{Tab:compare_NN_hardware}
\end{table}

\noindent
a. The energy efficiency is quoted 346 fJ/OP, which corresponds to 2.9 TeraOP/J.\\
b. The throughput in Ref.~\cite{Ashtiani2022} is 0.27 TeraOP/s with a chip area of 9 mm$^2$, corresponding to a throughtput density of 0.03 TeraOP/(mm$^2$$\cdot$s).\\
c. The end-to-end energy efficiency in Ref.~\cite{Ashtiani2022} is 14 pJ/OP, which is inverted to 0.07 TeraOP/J.\\
d. Optical energy consumption includes the power used for laser generation and driving EO modulators for data encoding.\\ e. Full system performance includes optical energy consumption, nonlinear activation, data readout, signal amplification, ADC, DAC, memory access. 
\clearpage

\section{Fanout of Weights} \label{Sec:weights_fanout}

\begin{figure*}[ht]
    \centering
  \includegraphics[width=180mm]{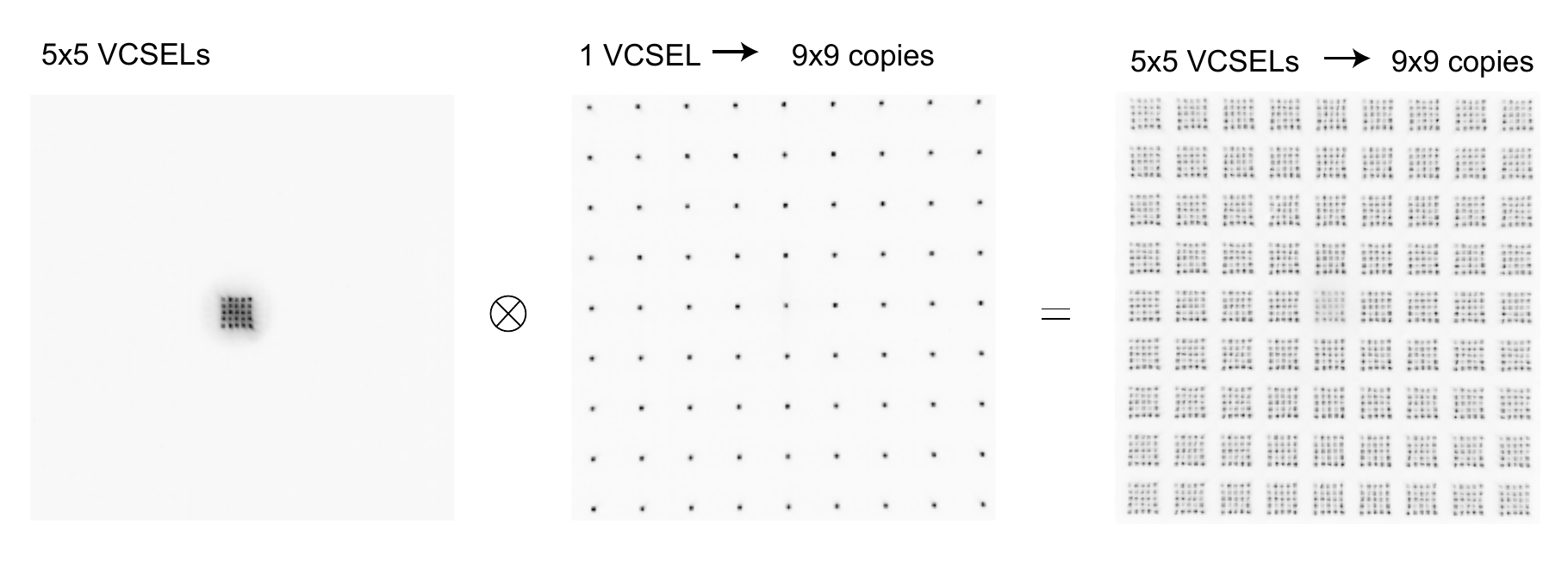}
    \caption{The 5$\times$5 VCSELs in a array is fanned out to k=9$\times$9 copies, potentially enabling to process 81 input vectors simultaneously.}
    \label{fig:fanout}
 
\end{figure*}

In the proposed scheme, the weight beams that encode the weight matrix [W$_0$, W$_1$, ..., W$_j$] can be fan-out to $k$ copies, enabling to multiply with a batch of $k$ input vectors simultaneously. Figure \ref{fig:fanout} shows the beams from the 5$\times$5 VCSEL array on a camera. In the center plot, one of the VCSEL is fanned out with a phase mask (MS-225-970-Y-A, Holoor.co) that splits the beam to 9x9 copies. On the right, the 5$\times$5 VCSELs are lighted up, the whole VCSEL array is fanned out. The third image is mathematically equivalent to the convolution of the two images on the left.

\clearpage

\clearpage 
\section{VCSEL arrays} \label{sec:VCSEL_arrays}

\begin{figure*}[ht]
    \centering
    \includegraphics[width=160mm]{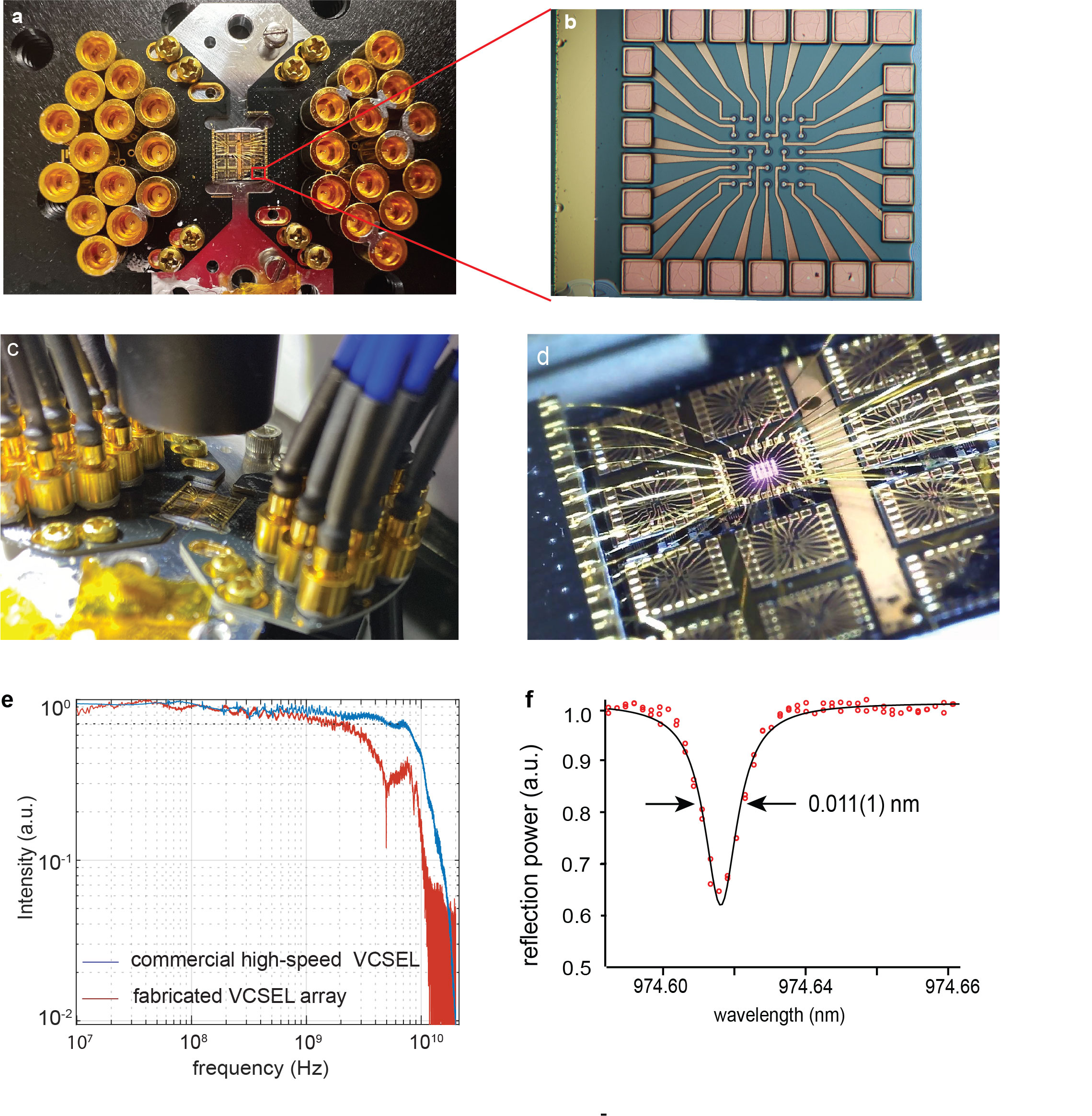}
    \caption{VCSEL samples. \textbf{a.} wire-bonded VCSEL arrays. The sample consists of 16 arrays, each with 5$\times$5 VCSELs. The VCSELs are individually wired to an external co-axis cable using a PCB. \textbf{d.} High resolution image with 5$\times$5 VCSELs. \textbf{c.} The sample in the setup. The beams are collected with an acromatic lens. \textbf{d.} 5$\times$5 wire-bonded VCSELs emits simultaneously.  \textbf{e.} the 3 dB bandwidth of the fabricated VCSEL is 2 GHz. The bandwidth of a commercial VCSEL is measured 7 GHz. \textbf{f.} the linewidth of the on-chip VCSEL resonance is 0.011(1) nm, corresponding to a cavity Q factor of $10^5$.}
    \label{fig:VCSEL_packaging}
\end{figure*}

One of our fabricated samples consisting of 16 arrays of $5\times5$ VCSELs (with 400 VCSELs in total) is shown in Fig.~S\ref{fig:VCSEL_packaging}.  We wire bond the arrays to a high-speed printed circuit board (PCB), which connects each individual VCSEL to an electronic driver using a bias tee. At the DC port of each bias tee, a single piece of AA battery (1.5 V) supplies all the VCSELs with a variable voltage divider on each channel. That allows individual voltage biasing to finely tune the wavelength of the VCSELs for injection locking. The AC port of the biased tee is connected to a high-speed arbitrary waveform generator (AWG) for data encoding. The output voltage of the AWGs is attenuated to 4 mV, matching to the $V_\pi$ of the VCSELs. 

The bandwidth of the VCSELs is measured with a vector network analyzer. The 3 dB bandwidth of our fabricated VCSEL is 2 GHz. We scanned the cavity resonance with an external tunable laser and measured the reflection spectrum. The Lorentzian fit to the the VCSEL resonance reveals a full-width-at-half-maximum of $\delta$$\nu$=0.011(1) nm, corresponding to 3 GHz. The Q-factor of the VCSEL cavity is Q=$\nu$/$\delta$$\nu$=10$^5$. The measured VCSEL bandwidth is 67~$\%$ of the photon-lifetime limit. The discrepancy might be due to the cavity dynamics modified by the laser emission (the linewidth scan is performed below lasing threshold, while the modulation bandwidth is measured above lasing threshold). To explore the potential of the VCSEL platform, we employ a commercial VCSEL \cite{VIS_VCSEL_1550} is about 7 GHz and it was used to demonstrate the data encoding at 25 GS/s in the main text.

\clearpage

\section{Injection locking} \label{sec: Injection-locking}

\begin{figure*}[ht]
    \centering
    \includegraphics[width=140mm]{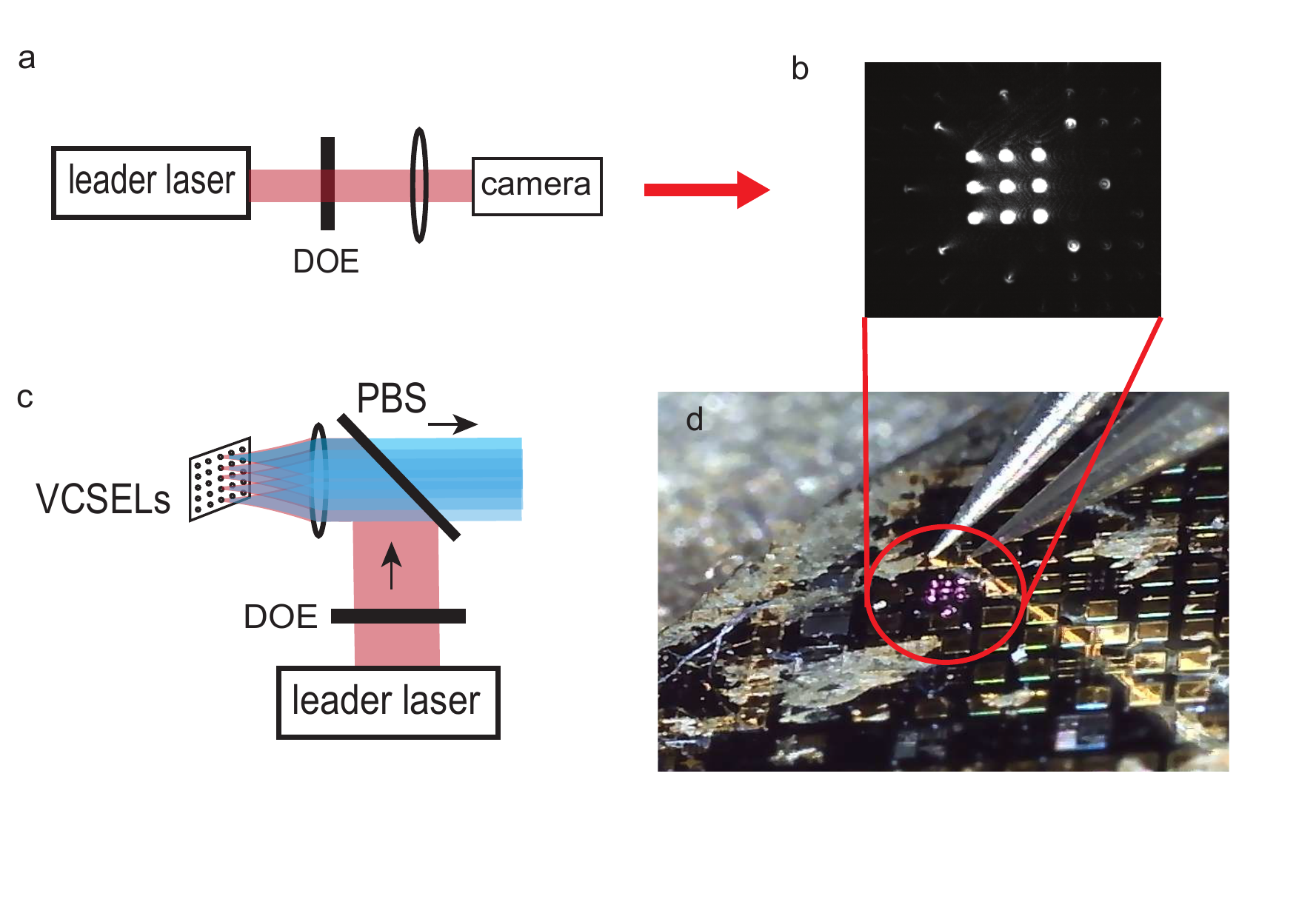}
    \caption{Example of injection locking over a VCSEL array. DOE: diffractive optical element. PBS: polarizing beam splitter. }
    \label{fig:injection_locking}
\end{figure*}

The setup of injection-locking is shown in Fig.~S\ref{fig:injection_locking}. Here the sample is with arrays of 3$\times$3 VCSELs. In \textbf{a}, the leader laser passes through a diffractive optical element (DOE, model MS-711-K-Y-X, HOLOOR.co) is fanned out to a 3$\times$3 beam array with separation angle of 0.026 degrees in the Fourier plane (Fig. ~S\ref{fig:injection_locking}b). The focus length of the lens is chosen 17.5 mm to provide a separation distance of 80 $\mu m$ between the beam spots, matching to the pitch of our VCSEL sample Fig.~S\ref{fig:injection_locking}d. For injection locking a 5$\times$5 array, the leader laser is fanned out to 5$\times$5 using another DOE (MS-844-K-Y-X, HOLOOR.co) with seperation angle of 0.018$\times$0.018.

\begin{figure}
    \centering
      \includegraphics[width=180mm]{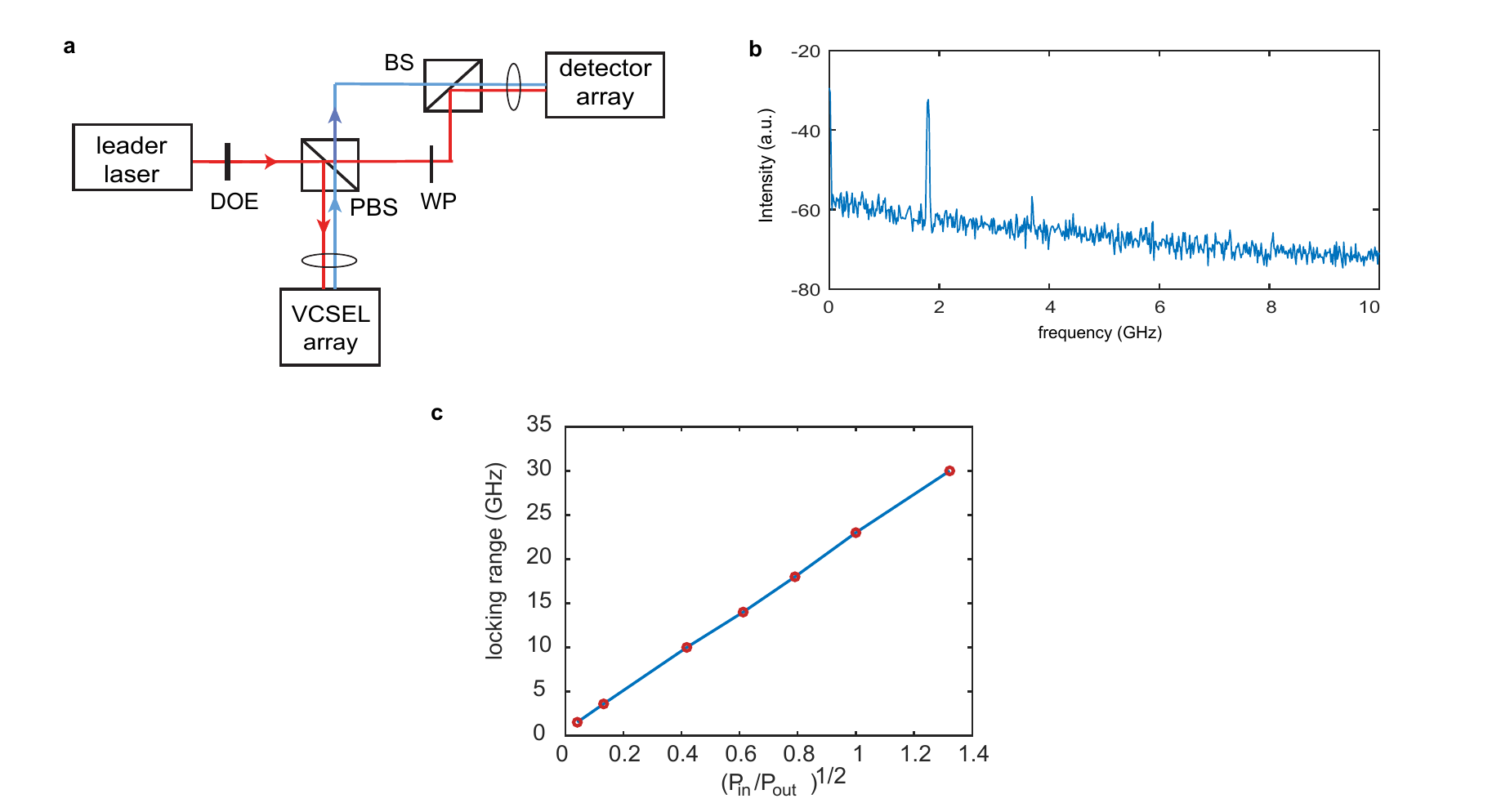}
    \caption{Injection locking range. \textbf{a.} experimental setup for characterization. DOE: diffractive optical element. PBS: polarising beam splitter. WP: half waveplate. BS: beam splitter. \textbf{b.} beatnote of the leader laser and a VCSEL at 1.9 GHz, out of injection lock range. \textbf{c} the injection lock range is proportional to the square root of the input laser power. The VCSEL output power is fixed 100 $\mu$W.}
    \label{fig:injection_locking_range}
\end{figure}

We perform inteferometric detection to characterize the injection locking of the VCSEL array with an experimental setup in Fig.~S\ref{fig:injection_locking_range}. The leader laser passing through the DOE is split to two parts by the polarizing beam splitter. One part is used for injection locking (Fig.~S\ref{fig:injection_locking}), and the other part passing is overlapped to the beams of the VCSELs using a beamsplitter. We couple the combined beams (one VCSEL and a copy of the leader laser) to a 4$\times$4 fiber array of 160-mm pitch (twice as that of the VCSELs) with a focus lens of 35-mm focus length.  The beatnote between the leader laser and a VCSEL is detected, as shown in Fig.~\ref{fig:injection_locking_range}b. We measured the injection locking range as a function of injection power in Fig.~\ref{fig:injection_locking_range}c by monitoring the beatnote with a high-speed photodetector (bandwidth 12 GHz) and a spectrim analyzer. The beatnote shifts to DC frequency when the VCSEL falls within the injection locking range. We tuned frequency of the leader laser while reading out the detuning using a wavemeter. The observed injection locking range is proportional to the square root of the input power. 

\begin{figure*}[!t]
    \centering
    \includegraphics[width=180mm]{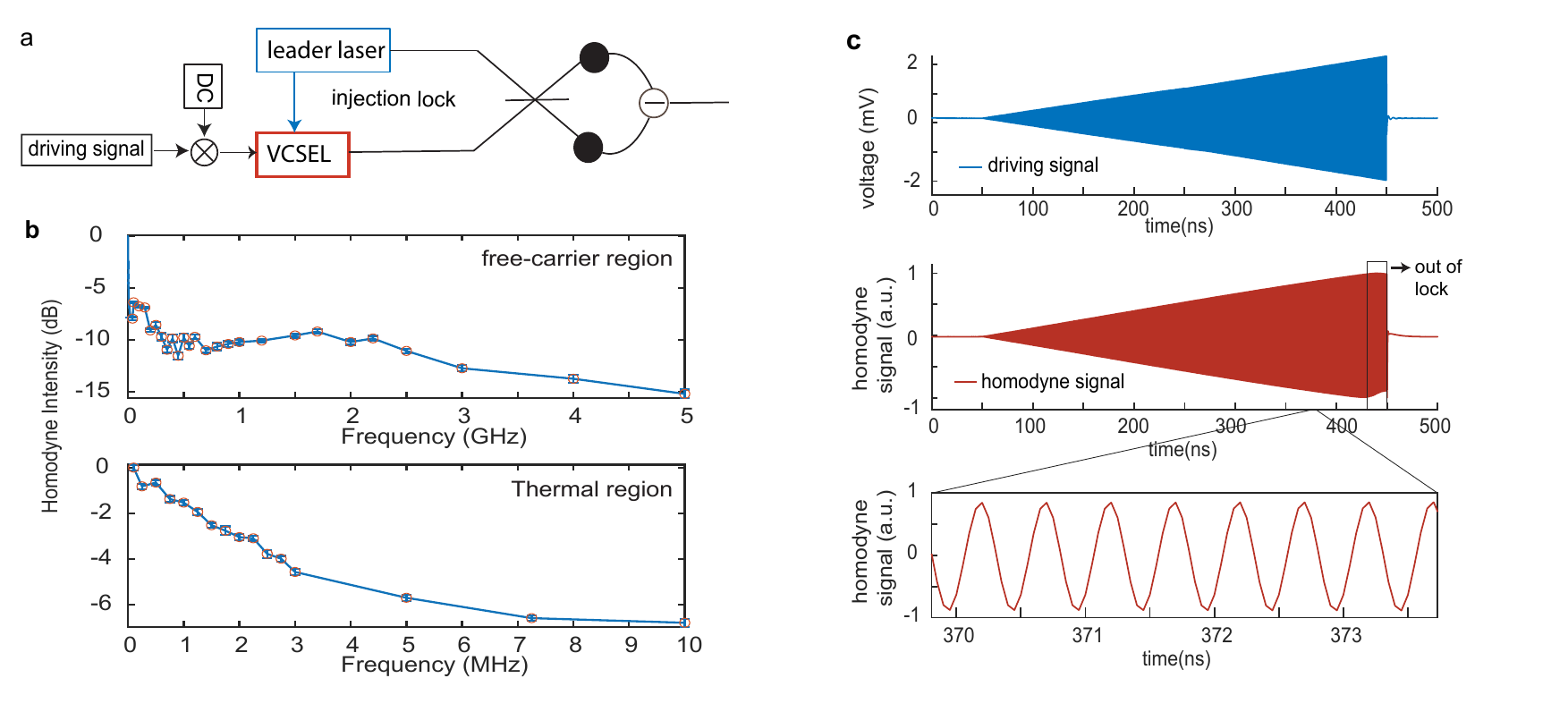}
    \caption{Homodyne detection of injection locking bandwidth and $V_\pi$. \textbf{a.} Setup of homodyne detection. \textbf{b.} frequency response of injection locked VCSEL. \textbf{c.} experimental characterization of $\pi$ phase shift voltage at 2 GHz.}
    \label{fig:injection_locking_Vpi}
\end{figure*}

We characterize the frequency response of the injection-clocked VCSEL with homodyne balanced detection (Fig. \ref{fig:injection_locking_Vpi}). The leader laser and the injection-locked VCSEL is overlapped with a beamsplitter and a balance detector records the homodyne interference. By applying a sine wave increasing frequencies (and constant amplitude), we observe an amplitude decay in the homodyne signal as a function of frequency (\ref{fig:injection_locking_Vpi}b). The result reveals that the frequency response of our injection-lock VCSELs in the free-carrier region (f>10 MHz) is 10 times weaker than the thermal region (f<1 MHz). Similar frequency response has been observed in Ref. \cite{Bhooplapur:11, phase_modulation_VCSEL}.

The $\pi$ phase shift voltage is measured by applying a sine wave with increasing voltage using the data encoding scheme of (Fig.~S\ref{fig:injection_locking_Vpi}a). As shown in Fig.~S\ref{fig:injection_locking_Vpi}c, the driving voltage (blue) increases from 0 to 4.2 mV (in x-axis from 50 ns to 450 ns). The homodyne signal increases linearly with the driving voltage (red) until the VCSEL falls out of injection lock at the peak-to-peak voltage of about 4 mV, which suggests a $V_\pi$=4 mV. The envelope of this homodyne signal is plotted as a function of driving voltages in the main text Fig. 2h.

\clearpage
\section{Data modulation and demodulation}

\begin{figure*}[ht]
    \centering
  \includegraphics[width=180mm]{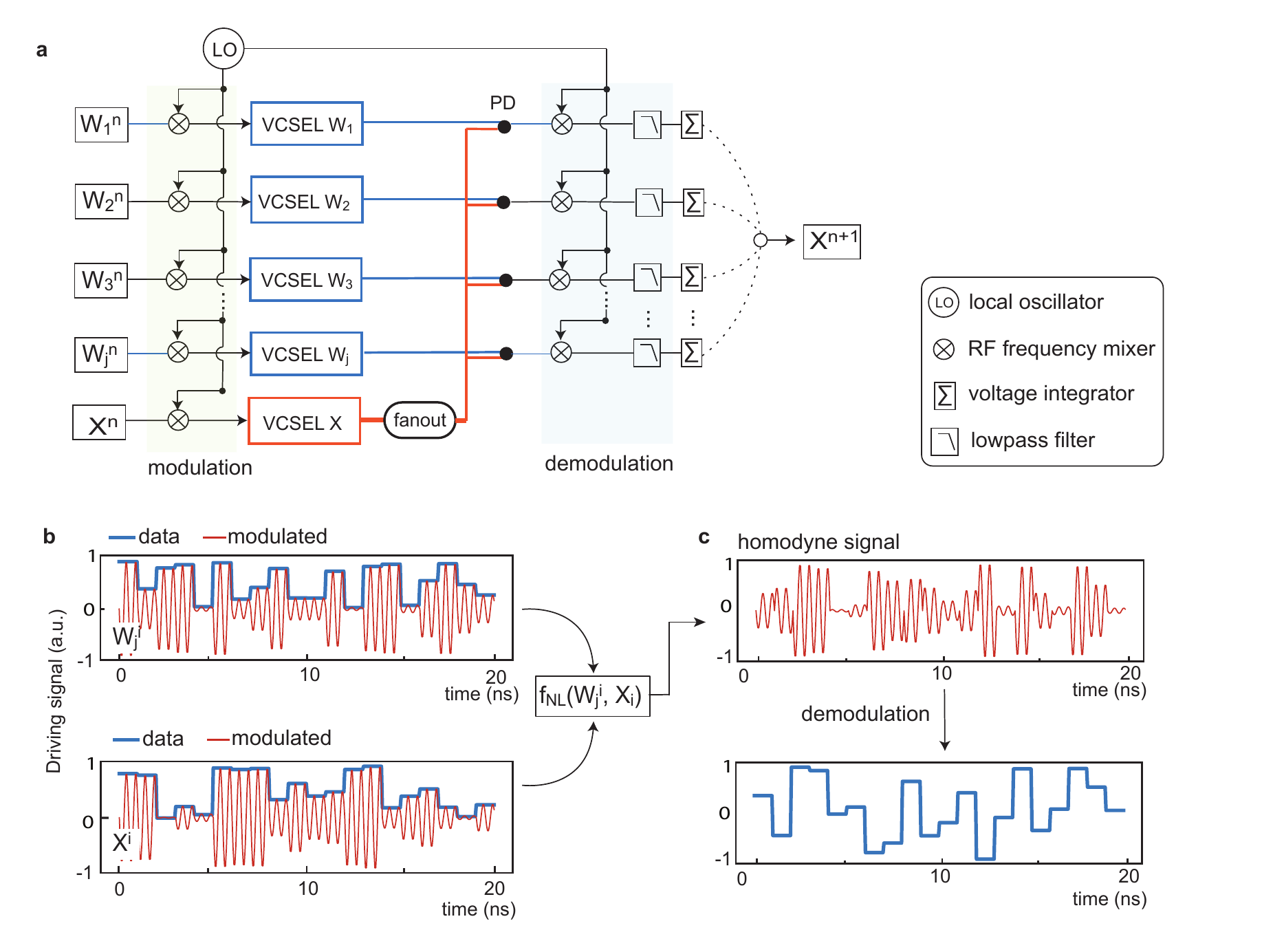}
    \caption{Data modulation and demodulation. (a) The data modulation scheme in the experimental setup. A local oscillator is used for data modulation and demodulation. LO: local oscillator. PD: photo-detector. \textbf{b.} Modulated data. the input and weight data are encoded by mixing with a local oscillator at the frequency of two times of the datarate. \textbf{c.} The generated homodyne signal is demodulated with the local-oscillator to retrieve the compute result. }
    \label{fig:data_modulation}
\end{figure*}

The thermal response on our injection-locked VCSELs leads to slow phase drifts in the homodyne signal at 1 GS/s speed modulation. To decouple the thermal effect, we encode the data (at 1 GS/s) with a fast local oscillator at $\omega_{\textsc{LO}}=2\pi\cdot$2 GHz using a frequency mixer, as shown in Fig.~S\ref{fig:data_modulation}. 
The modulated driving signal (Fig.~\ref{fig:data_modulation}b) averages at 0, which is good for thermal balance in every time step. The generated homodyne signal is represented as

\vspace{-7mm}
\begin{equation}\label{eq:data-modulation}
f_{NL1}(W_{ij}, X_i)=W_{ij}\sin(\omega_{LO})\sqrt{1-X_i^2\sin^2(\omega_{LO})}-X_i\sin(\omega_{LO})\sqrt{1-W_{ij}^2\sin^2(\omega_{LO})}
\end{equation}

\begin{figure*}[ht]
    \centering
  \includegraphics[width=180mm]{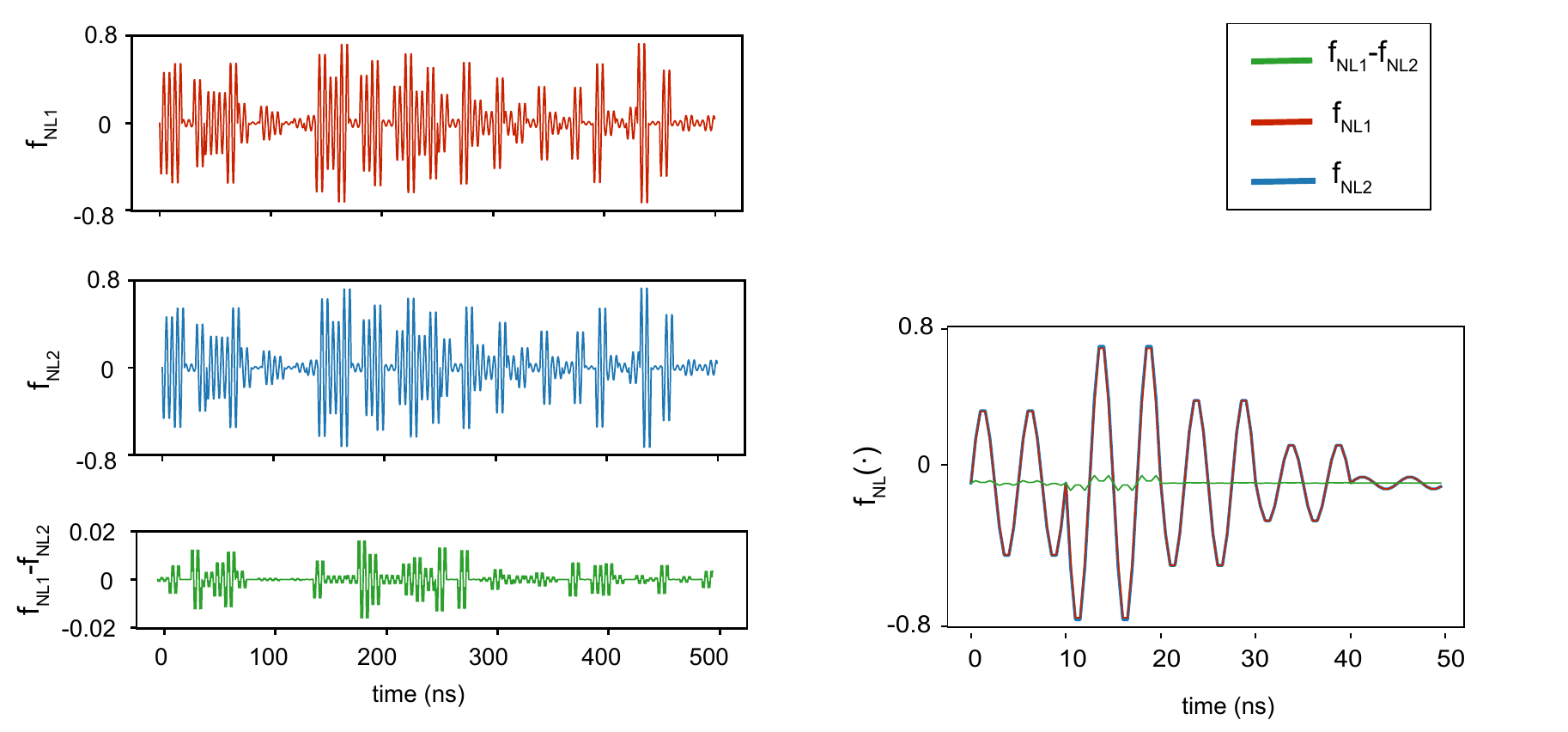}
    \caption{ Verification of compute fidelity using data modulation scheme. The red curve is computed with $f_{NL1}$ and the blue is from $f_{NL2}$. The $f_{NL1}-f_{NL2}$ residual is 0.7~$\%$ over 10,000 data points, which is negligible.}
    \label{fig:data_modulation2}
\end{figure*}

To confirm that the signal can be demodulated correctly. We compare the simulation result of Eq. \ref{eq:data-modulation} to that of $f_\textsc{NL2}$ in Fig. S\ref{fig:data_modulation2}  
\vspace{-3mm}
\begin{equation}\label{eq:nonlinear_compute1}
f_{NL2}(W_{ij}, X_i)=\bigg[W_{ij}\sqrt{1-X_i^2}-X_i\sqrt{1-W_{ij}^2)}\bigg]\sin(\omega_{LO}),
\end{equation}

 where the sine wave can be demodulated to produce accurate homodyne product. 

\clearpage

\bibliography{supplemental.bib}{}